\documentclass[10pt,journal]{IEEEtran}
\IEEEoverridecommandlockouts
\usepackage{amsfonts}
\usepackage{multirow}
\usepackage{amsmath,amssymb}
\usepackage{amsmath}
\usepackage[normalem]{ulem}
\usepackage{algorithm}
\usepackage{graphicx}
\usepackage{color}
\usepackage{booktabs}
\usepackage{algpseudocode}
\usepackage{enumitem,kantlipsum}
\usepackage{gensymb}
\usepackage{todonotes}
\usepackage{amssymb}
\usepackage[detect-none]{siunitx}
\usepackage{soul}
\usepackage{caption}
\usepackage{lipsum}
\usepackage{tabularray}
\usepackage[export]{adjustbox}
\usepackage{pgfplots}
\usepackage{pgfplotstable}
\usepackage{graphics}
\usepgfplotslibrary{fillbetween}
\usetikzlibrary{pgfplots.groupplots}
\usetikzlibrary{patterns}
\usepackage[export]{adjustbox}

\usetikzlibrary{tikzmark, positioning, fit, shapes.misc}
\usetikzlibrary{decorations.pathreplacing, calc}
\tikzset{brace/.style={decorate, decoration={brace}},
    brace mirrored/.style={decorate,decoration={brace,mirror}},
}
\usetikzlibrary{shapes.geometric}
\pgfdeclareplotmark{mystarc}{
    \node[star,star point ratio=2.25,minimum size=6pt,
          inner sep=0pt,fill= black, draw = yellow!55!white] {};
}

\algdef{SE}[DOWHILE]{Do}{doWhile}{\algorithmicdo}[1]{\algorithmicwhile\ #1}%
\usepackage{xcolor}

\usepackage{hyperref}

\usepackage{tikz}

\newcommand{\secref}[1]{Sec.~\ref{#1}}
\newcommand{\figref}[1]{Fig.~\ref{#1}}
\newcommand{\tabref}[1]{Table~\ref{#1}}

\newcommand{\lone}{$\mathrm{L}_{\mathrm{1}}$ }
\newcommand{\ltwo}{$\mathrm{L}_{\mathrm{2}}$ }
\pgfplotsset{compat=1.18}

\begin{document}
\title{Parallelization is All System Identification Needs: End-to-end Vibration Diagnostics on a multi-core RISC-V edge device}
\author{Amirhossein~Kiamarzi,
        Amirhossein~Moallemi,~\IEEEmembership{Graduate Student Member,~IEEE,} Federica~Zonzini~\IEEEmembership{Member,~IEEE,}, Davide~Brunelli,~\IEEEmembership{Senior Member,~IEEE,}
        Davide~Rossi,~\IEEEmembership{Senior Member,~IEEE, }
         and~Giuseppe~Tagliavini,~\IEEEmembership{Member,~IEEE}

\IEEEcompsocitemizethanks{\IEEEcompsocthanksitem A. Kiamarzi, A. Moallemi, F. Zonzini, D. Brunelli, and D. Rossi are with the Department of Electrical, Electronic and Information Engineering, University of Bologna, 40136 Bologna, Italy\{\{amirhossein.kiamarz2, amirhossein.moallem2, federica.zonzini, davide.brunelli, davide.rossi\}@unibo.it\}.
\IEEEcompsocthanksitem G. Tagliavini is with the Department of Computer Science and Engineering, University of Bologna, 40136 Bologna, Italy \{giuseppe.tagliavini@unibo.it\}.
\IEEEcompsocthanksitem D. Brunelli is also with the Department of Industrial Engineering, University of Trento, 38123 Trento, Italy \{davide.brunelli@unitn.it\}.
} 
\thanks{This research was partially funded by PNRR – M4C2 – Inv. 1.3, PE00000013 – “FAIR” project – Spoke 8 “Pervasive AI,” funded by the EU under the NextGeneration EU programme. Further, this work was supported by the Italian Ministry for University and Research (MUR) under the program “Dipartimenti di Eccellenza (2023-2027)".} 
\thanks{The codes of this manuscript can be found at \url{github.com}}
}

\markboth{IEEE Internet of Things Journal,~Vol.~xx, No.~xx, February~2024}%
{Shell \MakeLowercase{\textit{et al.}}: Bare Demo of IEEEtran.cls for Computer Society Journals}

\IEEEtitleabstractindextext{%
\begin{abstract}
%

The early detection of structural malfunctions requires the installation of real-time monitoring systems ensuring continuous access to the damage-sensitive information; nevertheless, it can generate bottlenecks in terms of bandwidth and storage. Deploying data reduction techniques at the edge is recognized as a proficient solution to reduce the system's network traffic.
%
%
However, the most effective solutions currently employed for the purpose are typically based on memory and power-hungry algorithms, making their embedding on resource-constrained devices very challenging; this is the case of vibration data reduction based on System Identification (SysId) models. 
This paper presents PARSY-VDD, a fully optimized PArallel end-to-end \textcolor{black}{software} framework based on SYstem identification for Vibration-based Damage Detection, as a suitable solution to
perform damage detection at the edge in a time and energy-efficient manner, avoiding streaming raw data to the cloud.
First, we evaluate the damage detection capabilities of PARSY-VDD with two benchmarks: a bridge and a wind turbine blade, showcasing the robustness of the end-to-end approach.
Then, we deploy PARSY-VDD on both commercial single-core (STM32 family) and a specific multi-core (GAP9) edge device. 
\textcolor{black}{We introduce an architecture-agnostic algorithmic optimization for SysId, improving the execution by \qty{90}{\times} and reducing the consumption by \qty{85}{\times} compared with the state-of-the-art SysId implementation on GAP9.}
\textcolor{black}{Results show that by utilizing the unique parallel computing capabilities of GAP9, the execution time is \qty{751}{\micro \second} with the high-performance multi-core solution operating at \SI{370}{MHz} and \SI{0.8}{V}, while the energy consumption is \qty{37}{\micro J} with the low-power solution operating at \SI{240}{MHz} and \SI{0.65}{V}. Compared with other single-core implementations}
\textcolor{black}{based on STM32 microcontrollers}
\textcolor{black}{, the GAP9 high-performance configuration is \qty{76}{\times} faster, while the low-power configuration is \qty{360}{\times} more energy efficient.} 
%
%
\end{abstract}

}

\maketitle
\IEEEdisplaynontitleabstractindextext

\section{Introduction}
The omnipresent existence of complex structures is inseparable from human life. 
Indeed, engineered facilities, from long-span viaducts to residential buildings passing through industrial and aerospace components, serve as the backbone of modern society, easing transportation and improving the lifestyle~\cite{giordano2023value}. 
However, various external forces, such as earthquakes, extreme weather events, and aging, constantly challenge their integrity~\cite{iasha2020design, sohn2001damage}. 
Therefore, implementing real-time and long-lasting inspection strategies is imperative to ensure the safe exploitation of structures for human deployment, as malfunctioning can lead to irrecoverable human and budget loss~\cite{morgese2020post, figueiredo2013linear}. 

Structural Health Monitoring (SHM) systems specifically address this task thanks to the vertical coalescence between data acquisition, data transmission, and data analytics~\cite{dipietrangelo2023structural}.  
Nowadays, SHM architectures take advantage of distributed and very dense sensor networks, which are necessary to cope with the increasing dimension and complexity of the target assets~\cite{giordano2023value, parisi2022time, cai2016iot}.
Vibration-based approaches, in particular, are composed of densely deployed wireless accelerometer sensors and offer one of the most efficient solutions for the monitoring of structures in the dynamic regime, i.e., those which can be fully characterized by frequency-associated features~\cite{saidin2023vibration, kamariotis2023framework}.      

State-of-the-art vibration monitoring systems~\cite{polonelli2023self, di2021structural} exploit resourceful computational and transmission modules to stream data to an aggregation unit in the cloud in near real-time, using Internet-of-Things (IoT) technologies.    
The continuous flow of information generated by these networks introduces a considerable volume of data, raising two main issues.
One concerns the transmission and storage of such large volumes, which must be moved from the nodes to the cloud~\cite{cai2016iot}.
{For instance, considering a vibration sensor equipped with an 8-bit uni-axial accelerometer and programmed to sample 10 minutes of data at 100 Hz once per hour (according to typical vibration-based duty-cycles), a total amount of $\approx \SI{1.5}{MB}$ per day is generated; note that the same quantity scales unfavorably and easily exceeds tens of \si{G \byte} in case monitoring has to be performed continuously.}
%
{Secondly, transmitting such an extensive volume of data is power-demanding for battery-operated sensors based on wireless communication protocols, requiring frequent battery replacement. Among the tasks that a monitoring node has to perform, data outsourcing is one of the most energy-hungry operations. For example, if one considers narrowband IoT, which is one of the most modern protocols~\cite{ballerini2020nb}, a lifetime inferior to one year is estimated for a typical SHM sensor node performing six sessions of 10 minutes of monitoring per day \cite{di2021structural}. }
%

%
To address the former bottlenecks, {near-sensor data processing} has emerged {as a suitable means to reduce the size of the node-cloud data. Indeed, from such a perspective, only a pool of representative features extracted from the sensor itself must be transmitted in place of the raw time series, which is beneficial for the entire SHM system.} 
%
%
To this end, an unconventional but promising method specifically tailored for vibration time series has been proposed in~\cite{zonzini2022system}. This novel approach, based on System Identification (SysId) techniques, demonstrated to achieve more than \SI{50}{\times} vibration data reduction, outperforming other strategies in the same application domain. Nonetheless, deploying SysId on resource-constrained devices is highly challenging due to its computational complexity and huge memory requirements. Although Zonzini \textit{et al.}~\cite{zonzini2022system} showed its feasibility for deployment on a single-core microcontroller unit (MCU), this first attempt is still limited by high latency ($>$\qty{2}{\minute}) and energy consumption ($\approx$ \qty{100}{\milli \joule}). Such quantities are barely acceptable for real-field applications for two main reasons. Firstly, having long computing times forces the sensing unit to work in one of the most demanding power modes necessary to maximize performance; ultimately, this implies an unavoidable reduction of the energy budget and, thus, of the battery life powering the sensor. Then, high latency does not allow for the implementation of real-time monitoring (processing can become even more cumbersome than mere data acquisition), a functionality that must be ensured under critical monitoring scenarios. 

\textcolor{black}{Recent work by Moallemi \textit{et al.} \cite{moallemi2023speeding} explored using parallel QR for SysId on the multi-core GAP9 MCU. Their method aims to reduce latency and energy by performing the QR decomposition of small-size data chunks that are further combined according to a binary processing scheme. Contrariwise, in our work, we apply parallel QR to the entire input matrix, changing the parallelization paradigm to more efficiently utilize the GAP9’s resources. This approach eliminates the extra overhead associated with matrix reconstruction in subsequent iterations by removing the iterative methods of \cite{moallemi2023speeding} and applying the QR factorization in a single step to the entire input matrix. Moreover, while the focus in \cite{moallemi2023speeding} is only on the SysId component, our approach provides a fully parallelized end-to-end processing flow, including SysId, data preparation, and feature extraction, making it a more complete and efficient computational pipeline for vibration-based damage detection on edge devices.}

{Traditional architectures for embedded computing employ MCUs (e.g., the STM32 processors), mostly available as single-core devices with limited performance. However, the demand for more powerful and energy-efficient solutions has driven the development of parallel ultra-low-power (PULP) architectures, such as GAP9 \cite{GAP9} from GreenWaves Technologies. These platforms reduce the supply voltage in the near-threshold region and introduce specialized units and Instruction Set Architecture (ISA) extensions \cite{rossi2021vega} to improve performance and energy efficiency. }

To overcome the previous time-demanding and power-hungry deployments of SysId, this work proposes an end-to-end parallelized version of SysId compatible with continuous SHM by exploiting the unique processing capabilities of GAP9. The main contributions are as follows: 
%
\begin{enumerate}
    \item We present the PArallel SYstem identification for Vibration Damage Detection (PARSY-VDD) \textcolor{black}{software} framework, an open-sourced and optimized end-to-end parallel implementation of output-only SysId algorithms to reduce the data network rate in IoT-based SHM systems and detect structural damages\footnote{{https://github.com/ah-kiamarzi/PARSY-VDD}}.     
    \item We present an evaluation of the suitability of PARSY-VDD for the diagnostics of two relevant SHM facilities, i.e., a bridge viaduct in Switzerland and an industrial wind turbine blade. We achieve successful damage identification and \textcolor{black}{a spectral divergence} below \qty{4.30E-3}{} when comparing with a golden model implemented on a laptop via built-in MATLAB functions. 
    \item We present a comprehensive analysis of parallelization techniques, energy efficiency, and performance optimizations employed to deploy PARSY-VDD on GAP9. \textcolor{black}{Thanks to advanced algorithmic and mapping optimizations, PARSY-VDD achieved a performance improvement of \qty{90}{\times} speed-up and almost \qty{85}{\times} energy reduction with respect to previous standalone SysId implementations on the same computing architecture.}
 
\end{enumerate} 
\textcolor{black}{
We achieve an execution time of \SI{751}{\micro \second} using $8$ cores at \SI{370}{\mega Hz} and $0.8$ V in the high-performance configuration (HP) when deploying PARSY-VDD on GAP9. In the low-power configuration (LP), using $8$ cores at \SI{240}{\mega Hz} and $0.65$ V, we achieve an energy consumption of \SI{37}{\micro J}. These results correspond to a {\qty{76}{\times}} performance speed-up and {\qty{360}{\times}} higher energy efficiency compared to commercial single-core STM32H7 and STM32F4 MCUs, respectively.}
The rest of this paper is organized as follows. 
\secref{sec:back_ground} provides theoretical requirements for the proposed damage detection framework.
Further, \secref{sec:software_res} provides a numerical analysis of the proposed framework, starting from an in-depth comparison of three QR decomposition methods. Further, this section provides an evaluation of different configurations of the PARSY-VDD.
%
Next, \secref{sec:methodology} details the implementations of the PARSY-VDD on the target multi-core device, describing the parallelization techniques used to achieve a microsecond range execution time. 
Then, \secref{sec:hardware_res} reports the performance of PARSY-VDD on several MCUs, showing the features of the multi-core vs. single-core devices. 
\secref{sec:stoa} study related works to the proposed implementation. 
Finally, \secref{sec:conc} conclude this work. 
 
\section{Hardware \& Software Preliminaries} \label{sec:back_ground}

This section describes the hardware and software background behind this work. 
Initially, the signal processing techniques at the basis of SysId are introduced, paving the way to the end-to-end damage detection approach. 
Then, two commercial families of near-sensor computing platforms are presented, one based on single-core MCUs and the other on emerging multi-core computing units.
\begin{figure*}[tb]
    \centering
    \includegraphics[width = 1\linewidth]{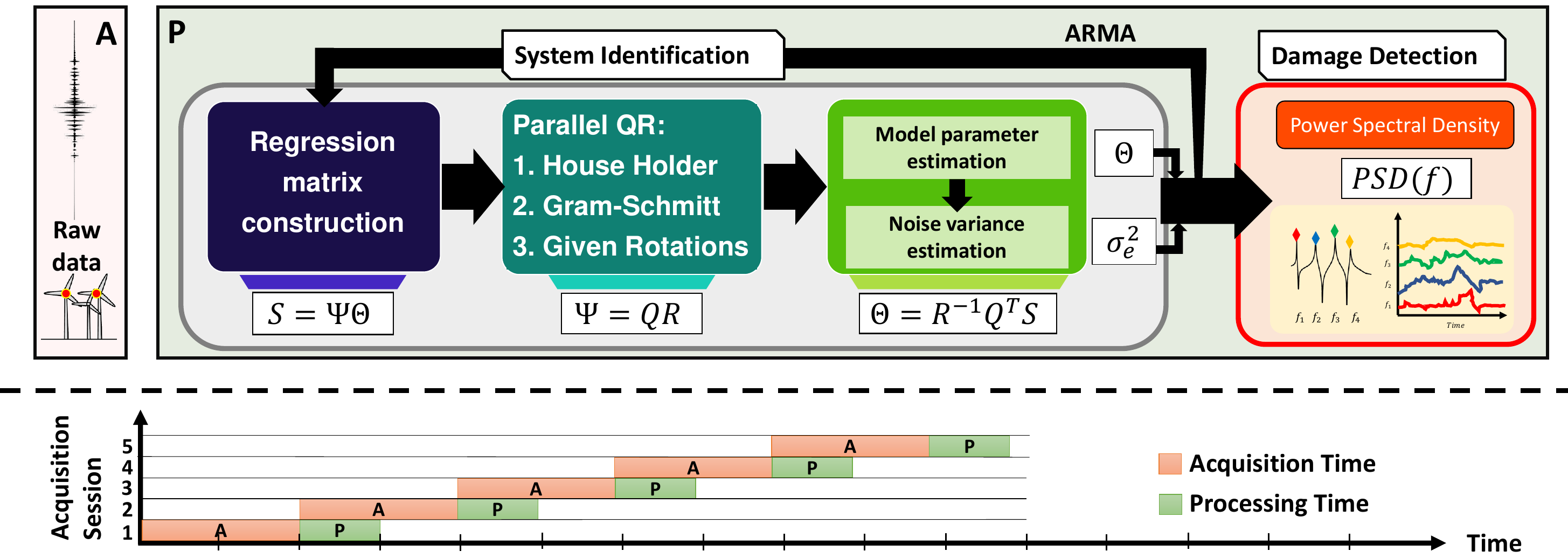}
    \caption{The main framework of PARallel SYstem identification for Vibration Damage Detection (PARSY-VDD). The bottom panel presents the acquisition and processing paradigm of the proposed framework, while the top part shows the data acquisition (block A), and the full end-to-end damage detection comprises SysId (either based on AR or ARMA, the latter requiring two iterations) and spectral-based damage detection (block P).}
    \label{fig:overview_anomaly}
\end{figure*}
\subsection{SysId based on Autoregressive models}\label{sec:sysid}
SysId refers to an ensemble of data-driven techniques postulating a mathematical model of a dynamic system from the measured output response. In vibration-based SHM, this translates into finding an equivalent mathematical abstraction of the structural behavior captured by vibration data (e.g., accelerations) whose characteristics could be further manipulated to retrieve representative damage indicators. 

SysId based on Autoregressive (AR) models implements such vision in an output-only manner, i.e., it is assumed that the output response is represented as a linear combination of its previous values plus a stochastic term. The latter is approximated as a Gaussian process having zero mean and variance $\sigma_e^2$, which serves as a proxy of the (not measurable and/or available) input stimulus. Noteworthy, output-only solutions are only doable for in-service SHM, where the structure is left to vibrate under the action of environmental and/or operational agents (such as vehicles, wind, etc.). 

Among the most effective AR models, the basic AR and its Moving Average (ARMA) variant, which differentiate in the complexity of the regression model, can be mentioned. Their mathematical formulations are expressed in Eq. (\ref{eq:y_ar}) and (\ref{eq:y_arma}), respectively:

    \begin{equation}\label{eq:y_ar}
        \begin{split}
            s[k] + \sum_{i = 0}^{q}\beta_i s[k-i] = e[k]
        \end{split}
    \end{equation}

    \begin{equation}\label{eq:y_arma}
        \begin{split}
            s[k] + \sum_{i = 1}^{q}\beta_i s[k-i] = e[k]+\sum_{s = 0}^{p} \alpha_s e[k-s]
        \end{split}
    \end{equation}
in which $s[k]$ and $e[k]$ are the output and input data samples at the generic discrete time $k$, correspondingly. Thereby, the objective of SysId is to estimate the noise variance $\sigma_e^2$ and the optimal set of $N_p = q+1$ coefficients $\Theta = [\beta_0 \dots \beta_q]$\footnote{$N_p = q+p+1$ and $\Theta = [\beta_0 \dots \beta_q \alpha_0 \dots \alpha_p]$ for the ARMA model.}, also called as \textit{model parameters}, that can best encapsulate the observed dynamics. Such a goal can be achieved via a four-step approach encompassing: 
\begin{enumerate}
    \item the creation of the regression matrix $\Psi$ corresponding to the chosen time series model, yielding to the regression form $S = \Psi \Theta, S \in \mathbb{R}^{[N \times N_p]} $;
    \item the decomposition of $\Psi$ via economy-size QR factorization, i.e., $\Psi = QR$, which returns the upper-triangular matrix $R \in \mathbb{R}^{[N_p \times N_p]}$ and the orthogonal matrix $Q \in \mathbb{R}^{[N \times N_p]}$;
    \item the computation of the sought parameters $\Theta = R^{-1}Q^TS$ via simple backward substitution (as enabled by the upper-triangular structure of $R$);
    \item the estimation of the noise variance according with $\sigma_e^2 = \frac{1}{N-N_p} ||S-\Psi\Theta||_2^2$.
    
    \end{enumerate}

While AR models follow this basic workflow, ARMA is more computationally involved since regression involves both the input and the output terms. Therefore, for ARMA, the procedure needs to be repeated twice: one stage is necessary to estimate $e[k]$ by fitting a high-order AR model to the time series, while the second step returns the actual model parameters. 
\subsection{Damage detection from model parameters} \label{sec:parsy}
The entire pipeline necessary to implement damage detection based on SysId is schematically depicted in \figref{fig:overview_anomaly} and comprises two main phases. The former is the SysId workflow described in \secref{sec:sysid}. 
The second step, instead, performs damage identification via a spectral-oriented approach leveraging the fact that natural frequencies are one of the most effective damage indicators~\cite{di2021structural}. 
More specifically, flaws can be identified as the appearance of relevant variations in the peak spectral values of the estimated spectral profile. 
This methodology has been selected since it is regarded as a standard practice in structural engineering, where model-based solutions are usually preferred over purely black-box approaches to preserve physical consistency in the damage-sensitive features~\cite{saidin2023vibration, jahangiri2023procedure}. However, we would like to specify that this spectral choice is nonbinding and, in principle, alternative pattern recognition techniques (either built on machine learning or standard statistics) are also applicable to address the same task (see, e.g., \cite{zonzini2023tiny}). 
Besides physical interpretation, another advantage of frequency-based detectors is that they require minimal (eventually null) effort in case adaptation is needed since they do not depend on fine-tuning critical hyperparameters.

Accordingly, the procedure hinges on the dual relationship between the time and frequency domain formulation at the basis of SysId. Particularly, it works by applying the Fourier transform to Eqs. (\ref{eq:y_ar}) and (\ref{eq:y_arma}) such that the power spectral density (PSD) of the signal can be estimated analytically from the sole knowledge of the model parameters and noise variance as

\begin{equation}\label{eq:psdarma}
    PSD_{AR}(f) =  \dfrac{\sigma_e^2} {\left |1+\sum_{i = 0}^{q} \beta_i e^{-j2\pi fiT_s}\right |^2}
\end{equation}

\begin{equation}\label{eq:psdar}
    PSD_{ARMA}(f) =  \sigma_e^2 \left | \dfrac{1 + \sum_{s = 0}^{p} \alpha_s e^{-j2\pi fsT_s}} {1+\sum_{i = 1}^{q} \beta_i e^{-j2\pi fiT_s}} \right |^2
\end{equation}
Notably, the spectrum can be obtained with arbitrary frequency resolution once the number of frequency points $L$ has been defined. This is one of the most relevant benefits of SysId concerning standard spectral analyzers, whose resolution usually depends on the number of points in the input signal. 

From the PSD profiles, the peak spectral values can be retrieved via post-processing algorithms. When their location and magnitude over time exceeds a safety threshold, an alarm can be issued.
\subsection{QR Decomposition} \label{subsec:qr_decomposition}
The \textit{QR} decomposition is a method that factors a generic \textit{m}-by-\textit{n} matrix \textit{A} into the product \textit{A = QR}, where \textit{Q} is an \textit{m}-by-\textit{m} unitary matrix, and \textit{R} is an \textit{m}-by-\textit{n} upper triangular matrix. In particular, QR represents the computing core of SysId; thus, its proper implementation deserves crucial attention. 

Various one-sided decomposition methods, including GR and HH~\cite{van1996matrix}, utilize orthogonal transformations to convert the input matrix into upper triangular form. By concatenating these orthogonal transforms, the matrix \textit{Q} is constructed while ensuring the invariants \textit{A = QR} and \(Q^H Q~=~I\) are preserved. Another approach, GS, achieves the {QR} decomposition through projection operations. Each of these methods possesses precise numerical properties and offers some degree of parallelism.

\subsubsection{Givens Rotations}
GR employs a sequence of plane rotations to nullify specific matrix elements. GR exhibits a low memory footprint and moderate computational complexity. Its efficacy extends to well-conditioned and ill-conditioned matrices, showcasing high stability for the latter.
\subsubsection{Gram-Schmidt}
GS operates orthogonal projections for matrix orthogonalization. Notably, the memory requirement is moderate, while the computational complexity is relatively low. Although GS demonstrates moderate stability for well-conditioned matrices, its susceptibility to numerical errors renders it less stable for ill-conditioned matrices.
\subsubsection{Householder}
HH employs orthogonal transformations through reflection vectors. This approach incurs a high memory footprint, primarily due to temporary storage requirements during matrix multiplication. With high computational complexity stemming from matrix multiplication, HH demonstrates high stability for well-conditioned matrices yet remains consistently stable for ill-conditioned matrices. To alleviate the computational complexity and high memory footprint inherent to the HH algorithm, an optimization strategy initially introduced in~\cite{GolubVanLoan2012} is also employed in this work. The approach involves utilizing the lower diagonal elements of the $R$ matrices to store the values of HH vectors, thus reducing memory requirements. Moreover, a backward accumulation technique is adopted to calculate the $Q$ matrices. While these strategies effectively alleviate the computational and memory overhead, the HH method maintains a higher computational complexity and memory footprint than the GR and GS approaches.
Based on these considerations, the choice of the matrix factorization method is contingent on matrix characteristics and the desired balance between memory usage, computational efficiency, and stability.

\subsection{{Commercial 32-bit MCUs}}
\subsubsection{STM32 Processors} \label{subsec:stm32}
The STM32 MCU processor family covers a wide range of single-core microcontrollers targeting three different criteria, i.e., high-performance with real-time constraints providing digital signal processing capabilities (\textit{F \& H Series}), ultra-low power consumption (\textit{L Series}), and wireless connectivity for IoT use-cases (\textit{W Series}).
STM32 MCUs are Arm\textregistered Cortex\textregistered-M-based devices with a 32-bit instruction set architecture (ISA)~\cite{yiu2013definitive}.
The nominal operation frequency of these devices varies from a few \qty{}{\mega \hertz} to a maximum of \qty{400}{\mega \hertz} depending on the series. 
They are mostly single-core devices featuring Real-Time-Operating free Systems (RTOS) to handle several tasks in near real-time. 
Next, they include a wide range of peripherals, making them suitable for embedded systems, enabling acquisition and processing at the edge.
The STM32 memory is limited to \qty{1.4}{\mega \byte} of Static RAM (SRAM) as the primary working memory to store variables and stack data. 
This technology does not require periodic refreshment to maintain its contents, unlike dynamic RAM (DRAM), and is faster than Flash memory.
In particular, \textit{H series} devices' SRAM is in the range of \qty{564}{\kilo \byte} to \qty{1.4}{\mega \byte}, while \textit{F series} covers less RAM with \qty{32}{\kilo \byte} to maximum \qty{256}{\kilo \byte}. Further, the Flash memory provided in each series varies between \qty{1} to \qty{2}{\mega \byte}, used to store code and make data persistent when the device is turned off.
{For the context of this work, we have used \textit{F \& H Series} to compare the single-core vs multi-core implementations of the PARSY-VDD configurations.}
\subsubsection{Greenwaves GAP9} \label{subsec_gap9_theory}

The GAP9 MCU~\cite{GAP9} is an ultra-low-power multi-core platform targeted for IoT applications at the edge. 
GAP9 features a single-core MCU-class core (fabric controller) orchestrating system-level operations (e.g., system boot and I/O connectivity) and a 9-core compute cluster to support parallel execution.
All the cores adhere to the RISC-V standard and support the PULP ISA extensions~\cite{rossi2021vega}.
The nominal frequency of the cluster can be increased up to \qty{370}{ MHz} while keeping the power consumption in the nominal operating mode below \qty{50}{\milli W}. 
Conversely, other multi-core RISCV-based MCUs or dual-core ARM Cortex devices provide fewer cores with the same or smaller clock frequency. 
GAP9 speeds up the execution of Digital Signal Processing (DSP) algorithms thanks to dedicated ISA extensions such as post-incremented Load (LD) and Store (ST), hardware loops, and packed Single Instruction Multiple Data (SIMD) instructions~\cite{Gautschi2017} and 4 dedicated floating-point units (FPUs) shared among the cores supporting 16-bit and 32-bit precision operations.
%
%

\begin{figure}[tb]
  \centering
\includegraphics[width=1\columnwidth]{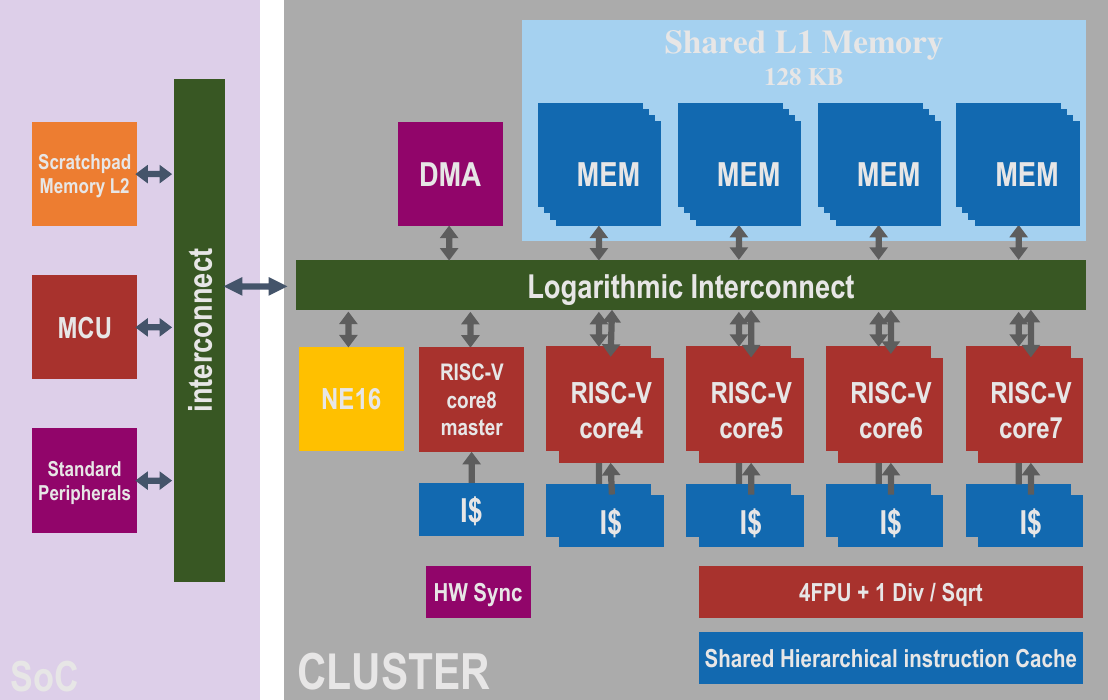}
  \caption{Block diagram of GAP9 Architecture.}
\label{fig:gap9}
\end{figure}

\figref{fig:gap9} depicts a simplified block diagram of the GAP9 architecture. It has a hierarchical memory architecture with \qty{128}{\kilo \byte} of single clock latency tightly coupled data memory, namely, \lone, as well as \ltwo SRAM with \qty{1.6}{\mega \byte} and an \ltwo non-volatile memory of \qty{2}{\mega \byte}.
{Moreover, \lone interconnect minimizes access contentions to the SRAM banks via a word-level interleaving scheme to evenly distribute the requests (upper part of the \figref{fig:gap9}).
Similar to memory, GAP9 also has a hierarchical program cache featuring 8 512-B private per core. 
Next, a joint combination of a parallel code and the \SI{4}{\kilo \byte} shared cache with two-cycle latency maximizes efficiency.
Finally, the \textit{event unit} embedded in GAP9 is a hardware unit accelerating the fine-grained parallel thread dispatching and synchronization, critical tasks for many applications exploiting single program multiple data parallelization schemes. Further, this unit is in charge of clock gating of the idle cores waiting for synchronization and enables resuming execution in two cycles~\cite{rossi2021vega}.}

\section{PARSY-VDD: from numerical to real-life applications} \label{sec:software_res}
In this Section, we focus on the performance assessment of the core blocks of PARSY-VDD, addressing the following aspects:

\begin{enumerate}
    \item We evaluate the accuracy of the three parallel QR methods in \secref{subsec:qr_decomposition} compared to built-in QR functions in MATLAB; this analysis is instrumental in determining the best QR solution in the context of the target SHM application;
    \item We extensively compare the two SysId models in terms of spectral fitting and damage detection capabilities, eventually identifying the most effective PARSY-VDD parameters.
\end{enumerate}
The employed SHM benchmark and the results obtained are presented below.

\subsection{Benchmarks}
Two benchmarks related to in-service facilities have been employed for the sake of PARSY-VDD structural performance assessment in both industrial and infrastructural scenarios. For each use case, three different signals have been processed, one for the healthy and two representative of unhealthy configurations. 
\subsubsection{Z24 Bridge} \label{subsubsec:z24}
The Z24 bridge was used to connect Bern to Z\"urich before its demolition, executed in 1999 for renovation. 
During the last year of service, the structure has been extensively monitored in different environmental and operative conditions, encompassing a phase of purposely induced defects. 
For our purposes, the nominal signal was selected at the beginning of the campaign when the structure was still in good condition. Then, a time series measured after lowering a pier and a one-time series related to the rupture of some tendons have been considered as indicative of medium and severe damage, correspondingly. 
\subsubsection{Sonkyo Wind turbine} \label{subsubsec:windturbine}
A small-scale sample of the Sonkyo Wind turbine blade, located in the experimental facilities of ETH Z\"urich, is another case study from the industrial field. 
It is characterized by a length of {\qty{1.75}{\meter}} and a nominal power of \qty{3.5}{\kilo\watt}. 
An accelerometer sensor network was installed on its surface during an experimental campaign~\cite{ou2021vibration} in which various types of defects and temperatures were simulated within a climate chamber.
The damaged samples for our dataset were collected when the wind turbine was experiencing one (medium) and two (severe) cracks in nominal temperature conditions.

\subsection{Analysis of QR decomposition methods}\label{sec:QR_analysis}
Before delving into the comparison of different QR decomposition methods, it is essential to define the quantitative metrics that will be used to evaluate the accuracy of each implementation. Such accuracy has been quantified in spectral terms by measuring the level of superimposition between the PSD profile returned by the GAP9 implementation and the one computed by a laptop device running QR algorithms in the MATLAB environment with double-precision floating point values. 
More specifically, we used the Itakura-Saito Spectral Divergence (ISD) 

\begin{equation}
    \small{ISD=\frac{1}{L}\sum_{f=1}^{L}}
    \left[
    \small{\frac{PSD_{GAP9}(f)}{PSD_{PC}(f)}}-\log
    \left(
    \small{\frac{PSD_{GAP9}(f)}{PSD_{PC}(f)}}
    \right)
    -1
    \right]
    \label{eq:ISD}
\end{equation}
which is an average measure of the point-wise spectral \textcolor{black}{divergence} between two power spectra, i.e., the profile computed by the parallel embedded version on GAP9 ($PSD_{GAP9}$) and the one, $PSD_{PC}$, estimated offline. ISD = 0 means perfect spectral superimposition, while values close to 1 indicate complete misalignment. The reason for such a metric is that $PSD_{GAP9}$ uniquely depends on the SysId parameters returned at the end of the QR decomposition; thus, ISD can also be used as an inverse means to judge the performance of the QR algorithms since eventual variations can only be attributed to numerical errors occurred during matrix decomposition, being identical the other PARSY-VDD components. 

In the analyses, three different lengths of the time series have been chosen, leading to three different SysId complexities and dimensions ([$N \times N_p$]) of the matrix to be decomposed: light ($[200 \times 8]$), medium ($[480 \times 16]$), and heavy ($[2520 \times 56]$). The last two have specifically been selected because they correspond to the largest dimension of the regression matrix that can be accommodated in the \lone memory and \ltwo memory, respectively. 
Alongside, $L=2048$ frequency data points (in 32-bit floating-point precision) have been selected for the PSD computation.
\begin{table}[tb]
\centering
\caption{Comparison of Matrix factorization methods for $N\times N_p$ input}

\begin{tabular}{c | c c c} 
 \hline
 \hline
 \textbf{Method} & \textbf{Memory} & \textbf{Complexity \textcolor{black}{$(\simeq)$}} & \textbf{Well/Ill Stab}\\ 
 \hline
 
 \multirow{2}{*}{GR} & Low & Mid & \multirow{2}{*}{Mid/High}   \\ & \textcolor{black}{$\scriptstyle 2NN_p+ 4N $}   & \textcolor{black}{$\scriptstyle \frac{(6N_pN^2 - N_p^3)}{3}$}&\\ [1ex]
 
 \hline
 \multirow{2}{*}{GS}     & Mid  & Low  & \multirow{2}{*}{Mid/Low}  \\ & \textcolor{black}{$\scriptstyle 2NN_p +N_p^2 $}  & \textcolor{black}{$\scriptstyle (N_pN^2)$}&\\[1ex]

 \hline
 \multirow{2}{*}{HH}      & High & High & \multirow{2}{*}{High/High}\\ & \textcolor{black}{$\scriptstyle N_p + N + 5NN_p + N^2 $} & \textcolor{black}{$\scriptstyle \frac{(6N_p^2N^2 + N^4 -4N_pN^3)}{12}$}&\\[1ex]
 \hline
 \hline
\end{tabular}
\label{table:Methods}
\end{table}
\color{black}

\begin{table*}[tb]
  \centering
  \caption{Comparison of different QR decomposition methods [ISD] with different configurations of AR and ARMA models varying in the [$N \times N_p$] dimension of the regression matrix, and bit precision.}
  \label{tab:QR_isd}
  \begin{adjustbox}{width=\linewidth}
  {
  \renewcommand{\arraystretch}{1.2}
  \begin{tabular}{c|c c c | c c c|| c c c| c c c }
   \multirow{3}{*}{\rotatebox{90}{\textbf{Method}}}                 & \multicolumn{6}{c||}{\textbf{64-bit}} & \multicolumn{6}{c}{\textbf{32-bit}}\\ \cline{2-13}
                  & \multicolumn{3}{c|}{\textbf{AR}} & \multicolumn{3}{c||}{\textbf{ARMA}} & \multicolumn{3}{c|}{\textbf{AR}} & \multicolumn{3}{c}{\textbf{ARMA}}\\ \cline{2-13}
      & $200 \times 8$& $480 \times 16$& $2520 \times 56$ & $200 \times 8$& $480 \times 16$& $2520 \times 56$ & $200 \times 8$& $480 \times 16$& $2520 \times 56$ & $200 \times 8$& $480 \times 16$& $2520 \times 56$ \\                   
                                                        \hline \hline
    \textbf{GR} & 1.19E-15 & 3.70E-16 & 1.19E-16 & 7.11E-14 & 3.21E-13 & 1.17E-13 & 5.57E-08 & 8.13E-07 & 6.84E-07 & 3.98E-05 & 7.93E-05 & 1.47E-05\\
    \textbf{GS} & 6.14E-17 & 6.62E-16 & 3.47E-16 & 5.90E-12 & 1.00E-11 & 1.18E-12 & 5.36E-07 & 1.54E-06 & 2.74E-06 & 6.11E-03 & \textbf{4.30E-03} & 7.54E-04\\
    \textbf{HH} & 8.52E-16 & 1.85E-16 & 5.00E-17 &1.36E-13 & 1.97E-13 & 3.97E-14 & 4.80E-08 & 1.18E-07 & 1.42E-07 & 4.16E-05 & 5.84E-05 & 2.17E-06\\
                                                        \hline \hline 
  \end{tabular}
  }
  \end{adjustbox}
\end{table*}

~\tabref{table:Methods} compares the three QR methods discussed in \secref{subsec:qr_decomposition} in terms of memory occupation, computational complexity, stability under well-conditioned matrices, and stability under ill-conditioned matrices. 
GR and GS exhibit a simpler structure and lower time complexity $(O(n^3))$ compared to HH. This feature translates to a reduced number of instructions and a decreased total number of cycles. Conversely, HH presents the highest time complexity $(O(n^4))$. In particular, a significant portion of HH instructions is devoted to matrix multiplication involving the HH matrix, $R$ matrix, and $Q$ matrix. This phase generates values for $R$ and $Q$ matrices in subsequent iterations, significantly impacting the algorithm's overall cycle count. The simplicity of the GS method contributes, instead, to a reduction between one and two orders of magnitude in the execution time.

The inherently sequential nature of GR limits its capacity to leverage parallel execution, setting it at a disadvantage compared to the other methods. 
On the other hand, the structure of GS and HH makes them more suitable for parallel execution. GS benefits from an even workload distribution across each core in every iteration, and a significant portion of HH involves matrix multiplication, which can be efficiently parallelized. However, GS is expected to use less energy due to its fewer instructions and excellent capability for parallel execution, resulting in quicker performance compared to HH. Despite HH's effective parallelization, its high computational complexity leads to greater energy consumption \cite{10.1145/3649153.3649210}.

Considering the memory limitations of IoT systems, it is crucial to address the memory footprint of QR decomposition methods. In SysId, we primarily deal with tall and skinny matrices where the number of rows is one to two orders of magnitude larger than the number of columns (input matrix is $m$-by-$n$ with $m=N$ and $n = N_p$, $N \gg N_p$). Expressing the ratio of $N$ to $N_p$ as $M$ ($N/N_p=M$), the memory consumption of GS and HH methods \textcolor{black}{in ~\tabref{table:Methods} comes to be $N_p \times (N_p+2MN_p)$ and $N_p \times (1 + M + 5MN_p + M^2N_p)$, respectively, with GS having a complexity which is approximately $M$ times lower than HH.}


In addition to these architectural aspects, the accuracy of each QR decomposition method must also be taken into account when selecting the most effective solution. ~\tabref{tab:QR_isd} presents the ISD values for 64-bit and 32-bit implementations of each QR decomposition method in PARSY-VDD, considering both AR and ARMA as time series models across various regression matrix sizes. A general observation is that, with AR, the ISD values are typically one to three orders of magnitude lower than those for ARMA. It is important to note that, although the ISD values for GS are higher than those for GR and HH, even in the worst-case scenario (ARMA with 32-bit float implementation), they remain within the range of $10^{-3}$ \textcolor{black}{(the worst case being \qty{4.30E-3}{} as magnified in bold fonts)}, which is deemed more than acceptable for the considered applications.
These numbers prove that even if negligible deviations (close to machine precision) appear when 64-bit precision is considered, proving almost perfect matching between the GAP9 and the offline implementation under the same bit-depth, the 32-bit variant on GAP9 can still work accurately with uttermost benefits in terms of memory reduction. 

Taking into account the memory footprint, energy consumption, and performance of various QR decomposition methods, GS stands out as a preferable choice for IoT platforms with constrained computation, energy, and memory budgets.

\subsection{PARSY-VDD damage detection capability}

\begin{table}[tb]
  \centering
  \caption{Performance of PARSY-VDD for different configurations of AR and ARMA models varying in the [$N \times N_p$] dimension of the regression matrix. The considered criteria include ISD, $\Delta F$ (percentage difference in the most energetic peak between the healthy and damaged case) for both the medium \textcolor{black}{(Med)} and severe \textcolor{black}{(Sev)} defect, and memory.}
  \label{tab:isd}
  {
  \begin{tabular}{c|l|c|c|c|r}
        & \multirow{2}{*}{\textbf{Model}} & \textbf{ISD} & \multicolumn{2}{c|}{$\Delta F$ [\%]} & \textbf{Memory} \\ 
        & & [a.u.] & Med & Sev & [\si{\kilo \byte}]\\
        \hline
        \hline
        \multirow{6}{*}{\rotatebox{90}{\textbf{Wind turbine}}} & AR: $[200 \times 8]$ & 1.33E-06 & 2.25 & 4.10 & 13.888 \\ \cline{2-6}
        & AR: $[480 \times 16]$ & 1.63E-06 & 1.08 & 3.89 & 64.448 \\ \cline{2-6}
        & AR: $[2520 \times 56]$ & 9.19E-07 & 1.21 & 3.88 & 1151.808 \\ \cline{2-6}
        & ARMA: $[200 \times 8]$ & 1.91E-04 & 2.02 & 34.05 & 13.888 \\ \cline{2-6}
        & ARMA: $[480 \times 16]$ & 5.16E-04 & 1.35 & 4.54 & 64.448 \\ \cline{2-6}
        &  ARMA: $[2520 \times 56]$ & 8.72E-05 & 0.13 & 4.40 & 1151.808 \\ \hline
        \hline
        \multirow{6}{*}{\rotatebox{90}{\textbf{Z24}}} & AR: $[200 \times 8]$ & 9.20E-09 & 3.19 & 0.77 & 13.888 \\ \cline{2-6}
        & AR: $[480 \times 16]$ & 1.20E-07 & 1.16 & 0.39 & 64.448 \\ \cline{2-6}
        & AR: $[2520 \times 56]$ & 1.13E-06 & 0.26 & 0.13 & 1151.808 \\ \cline{2-6}
        & ARMA: $[200 \times 8]$ & 1.10E-07 & 0.56 & 5.84 & 13.888 \\ \cline{2-6}
        & ARMA: $[480 \times 16]$ & 1.51E-05 & 3.11 & 2.59 & 64.448 \\ \cline{2-6}
        & ARMA: $[2520 \times 56]$ & 1.36E-05 & 1.32 & 6.60 & 1151.808 \\ \hline
    \end{tabular}
  }
\end{table}

A comparison between ARMA and AR was first performed to evaluate their suitability for damage detection in the context of PARSY-VDD. To this end, the ISD with respect to the golden model running in MATLAB was first measured. Given the previous analysis, full parallelization for the GAP9 device (8 cores active) is assumed along with GS as the QR method, spanning the same model complexities already used in Sec. \ref{sec:QR_analysis}. 

Results are reported in \tabref{tab:isd} under the column `ISD', including the Z24 and the wind turbine use case.
As can be observed, all the values are stably below $1.91\cdot10^{\text{-}4}$ even for the worst-performing configuration, proving the spectral accuracy of the deployed solutions. 
Further, \tabref{tab:isd} reports that the performances for ARMA are two to three degrees of magnitude worse than AR's, in compliance with the QR outcome in Sec. \ref{sec:QR_analysis}. Furthermore, when increasing the matrix size to be decomposed (i.e., by increasing the number of parameters and length of the time series), no remarkable deterioration is noticed for AR, while a relevant improvement is observed for ARMA. This is implicitly due to the two-stage nature of the ARMA algorithm itself since the second stage of the QR decomposition is more ill-conditioned. 

\begin{figure*}
    \centering
    \includegraphics[width = \textwidth]{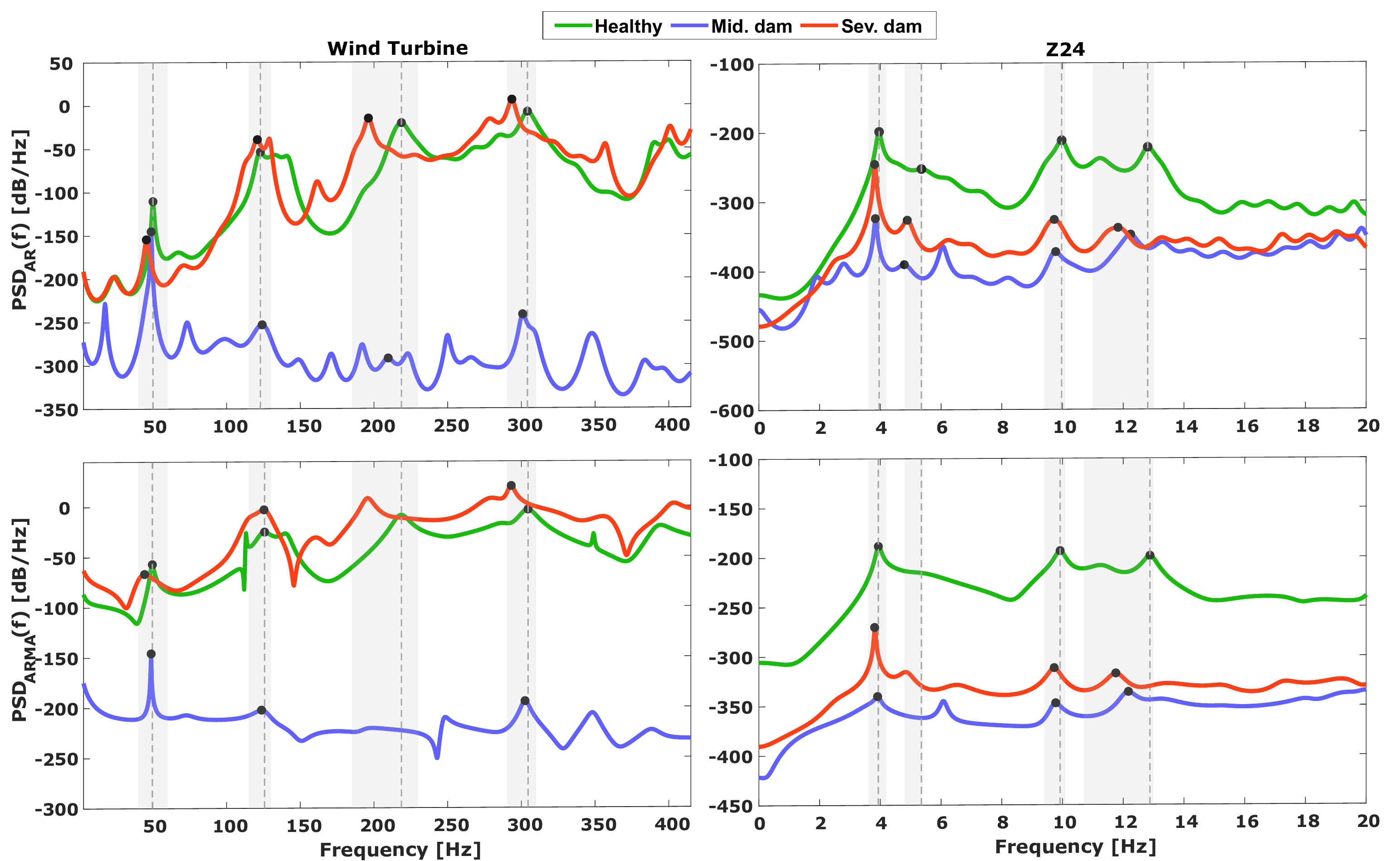}
    \caption{Spectra (AR - first row, ARMA - second row) for the wind turbine blade (left) and the Z24 bridge use case (right) obtained by the PARSY-VDD end-to-end application deployed at the extreme edge. In both cases, one signal in healthy conditions (green profiles) and two in differently damaged conditions (blue and red for mid and highly severe, respectively) are considered.}
    \label{fig:spectra}
\end{figure*}

Then, the applicability of the SysId models for damage detection has been investigated. PSD profiles for AR (first line) and ARMA (second line) with $N_p = 56$ are depicted in \figref{fig:spectra} for the wind turbine component (left plot) and the Z24 bridge (right panel). Three colors are adopted to indicate different structural configurations: green for the healthy condition and blue and red for the mid and severe damage status, respectively. 

Noteworthy, ARMA models allow the retrieval of comparatively smoother spectral profiles equal to the number of model parameters, thanks to the moving average term in the regression form. The proof is the fact that the spectra returned by ARMA contain much less spurious components than the AR ones, i.e., they are characterized by enhanced resolution and visible peak components. Such behavior is particularly evident for the Z24 bridge structure, while the differences are less pronounced for the wind turbine asset. Eventually, ARMA-based approaches allow the filtering out of the presence of false positives, and this is one of the main reasons motivating their adoption of AR in view of damage detection.

Despite these qualitative differences, it is worth noticing that the PARSY-VDD workflow can capture both subtle and more consistent variations in the spectral profile in all the considered configurations. 
This outcome is magnified by grey rectangles indicating the spectral bands in which the main peaks, namely, the damage-sensitive features, are located. 
In particular, it is possible to correctly track changes in the shape, width, and amplitude of the corresponding lobes, which can be interpreted as a symptom of structural deterioration.
Indeed, the positions of the markers in the green signals, related to the nominal condition, tend to reduce as the defect size increases, consistently with the structural dynamics theory \cite{brincker2015introduction}. For the Z24 structure, the highest reduction is observed for the fourth mode located at around \SI{13}{Hz}, which undergoes a contraction of more than \SI{1.2}{Hz} between the healthy and unhealthy signals. Alongside, the second and third frequencies shrink by \qty{10.7}{\%} and \qty{4.75}{\%}, respectively. The less evident variation is experienced for the first and most energetic mode located at around \SI{4}{Hz}, for which the reduction is around \qty{1}{\%}.

Similar behaviors are observed for the wind turbine blade, as testified by the sharp shift of the peak values and the energy distribution of the severe damage status (red curve) with respect to the reference one in green. The deviation is even more pronounced for the medium entity damage; interestingly, one peak around the third frequency disappears in the most degraded configuration, most likely due to some damping effects caused by the damage itself.

{A more detailed quantitative analysis of the damage detection capabilities of PARSY-VDD is provided in Table \ref{tab:isd} under the column '$\Delta F [\%]$'. $\Delta F = 100(1-f_{def}/f_{safe})$ expresses the smallest percentage variation of the most energetic peak between the healthy ($f_{safe}$) and defective ($f_{def}$) status. The reason for such a metric is that the smaller the detected frequency shift, the better the level of structural insight, i.e., even faint flaws can successfully be identified. }

{Results reveal two main outcomes. The former is that the higher the model complexity, the better the peak identification, i.e., smaller frequency variations can be tracked independently from the SysId model. Nevertheless, increasing $N_p$ implies a quadratic increment in the overall memory required to run the entire PARSY-VDD (\SI{13.8}{\kilo \byte} vs \SI{1151.8}{\kilo \byte} when comparing the lightest and heaviest configuration), confirming the importance of the appropriate selection of the SysId model configuration depending on the best compromise between performances and accuracy. The latter observation confirms that ARMA models usually perform more accurately compared to their AR counterparts.}
These observations hold independently from the target facility and the magnitude of the damage (the two assets experience comparatively different deterioration processes), corroborating the potential of the proposed approach for SHM purposes. 

\section{PARSY-VDD Deployment on GAP9}\label{sec:methodology}

\begin{figure}[t!]
\centerline{\includegraphics[width=0.5\textwidth]{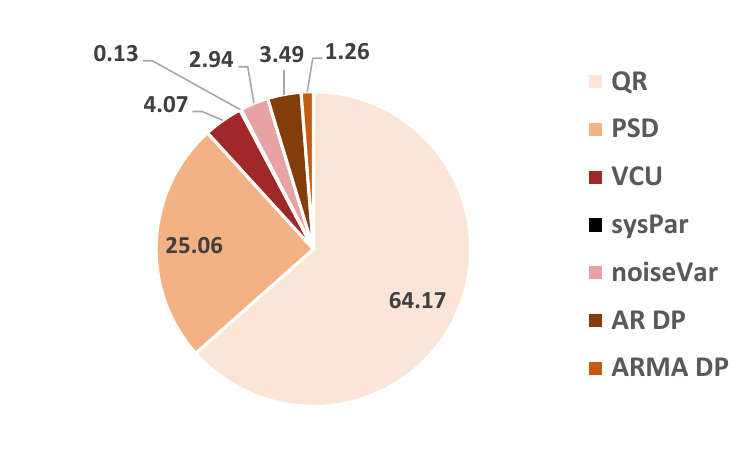}}
\caption{Code distribution of PARSY-VDD components. Abbreviations = 'DP': data preparation (construction of the regression matrix). 'VCU': vector coefficient update procedure (i.e., $Q^T S$), 'SysIdParam': estimation of the model parameters $\Theta$, and 'noiseVar': noise variance estimation.}
\label{figure:code-dist}
\end{figure}

\figref{figure:code-dist} illustrates the pre-parallelization distribution of code when utilizing ARMA models.
The largest portions of the code are dedicated to QR decomposition and PSD, accounting for over 89\% of the entire processing pipeline. In contrast, the code for model parameter estimation (SysPar) is the smallest share (only 0.13\%).
We designed a parallel version for all components (i.e., QR decomposition, PSD, Vector coefficient update, Noise variance estimation, and Data preparation for both AR and ARMA) except SysPar, whose contribution is negligible.
In the following paragraphs, we discuss the three aspects that required optimization on the GAP9 platform: data allocation, control flow parallelization, and code-level optimizations.

\begin{table*}[t]
\centering
\caption{\color{black}Performance analysis of PARSY-VDD.}
\begin{tabular}{c c c c c c c c} 
 \hline
 \hline
 \textbf{Cores}[\#] & \textbf{Tot. cycles}[\#] & \textbf{Core active}[\%] & \textbf{Ext. LD}[\%] & \textbf{LD-use stalls}[\%] & \textbf{L1 Stalls}[\%] & \textbf{I\$ miss stalls}[\%] & Barriers[\#]\\
 \hline
 
 1 & 1778862 & 99.92 & 0.01 & 0.00 & 2.45 & 10.32 & 0\\ 
                       2 & 923389  & 99.42 & 0.05 & 0.16 & 5.56 & 5.80  & 351\\
                      4 & 474995  & 98.67 & 0.09 & 0.86 & 5.57 & 7.83  & 351\\ 
                       8 & 277905  & 94.42 & 0.14 & 4.97 & 4.84 & 7.42  & 351\\ 
         \hline
         \hline
         
\end{tabular}
\captionsetup{width=1\linewidth}
\label{table:perf counter}
\end{table*}

%
%

\subsubsection{Data allocation}
\begin{figure}
    \centering
    \includegraphics[width = 0.5\textwidth]{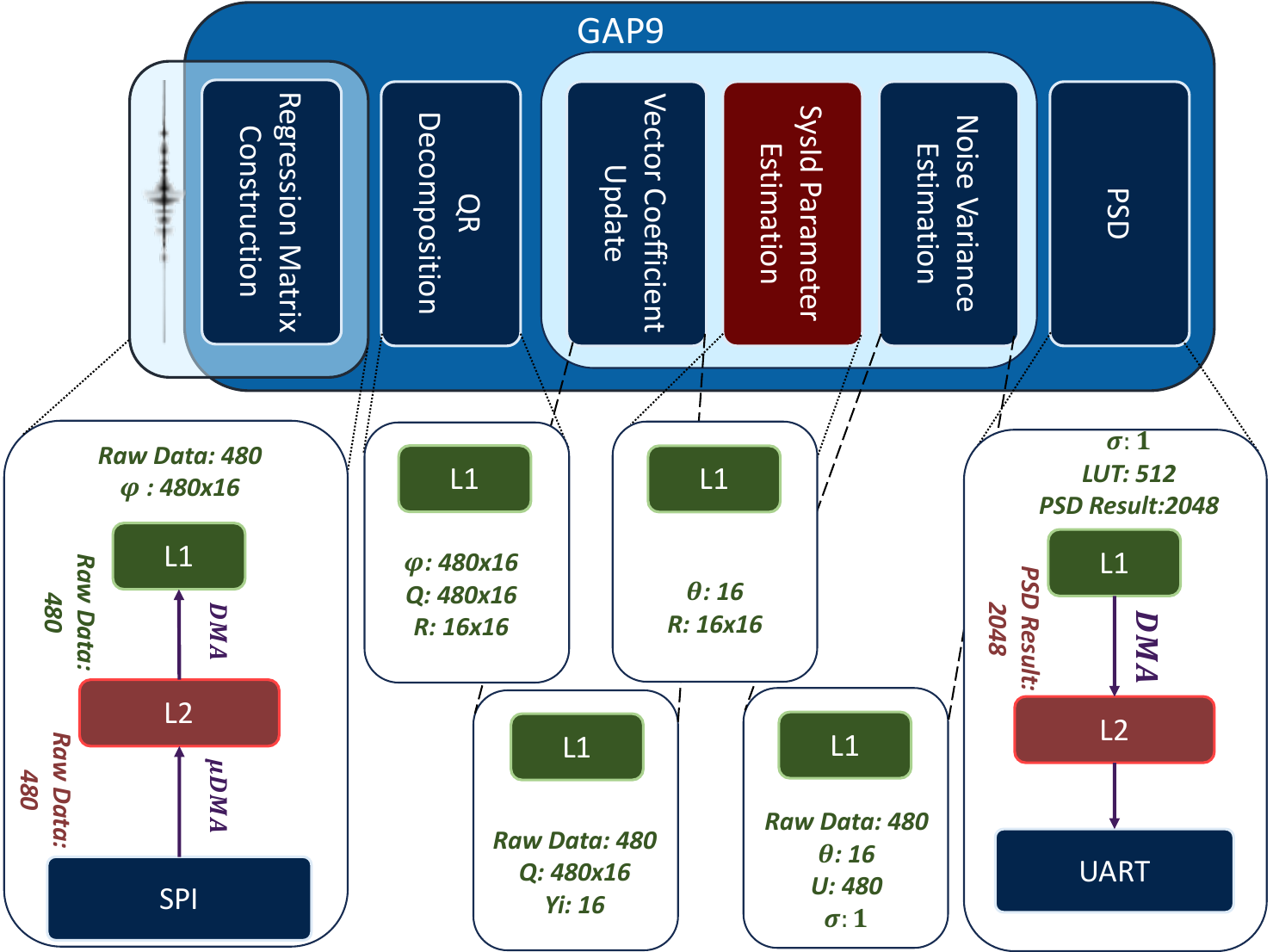}
    \caption{Processing pipeline and data allocation of PARSY-VDD components on GAP9. The Blue components are parallelized, while the red one (SysId parameter estimation) remains sequential.}
    \label{fig:GAP9-mem}
\end{figure}
Accessing \lone data requires a single-cycle latency for the cluster cores, while \ltwo access is ten times slower.
Accordingly, we assume to allocate the data structures within the \lone scratchpad memory to reduce memory access latency.
Figure~\ref{fig:GAP9-mem} illustrates the data allocation strategy for SysId. During the data acquisition and preparation stages, data is transferred from SPI to \ltwo memory using $\mu DMA$, and subsequently from \ltwo to \lone memory using DMA. The majority of computations occur in \lone memory, including the calculation of the PSD spectrum. Afterward, the results are moved back from \lone to \ltwo memory using DMA. 
\textcolor{black}{Given the DMA transfer speed of 8 bytes per cycle, two single-precision floating-point values can be loaded in each cycle. For data preparation, [$N - N_p$] input samples need to be transferred from L2 to L1 memory. In the case of $Np = 16$ and $N = 30Np$, using single-precision floating-point data (as specified in Section \ref{sec:hardware_res}), the data movement requires 232 cycles. For the PSD component, transferring 2048 single-precision floating-point samples requires 1024 cycles,}
which is negligible compared to the overall application computing cost.

\subsubsection{Control flow parallelization}

To exploit the GAP9 cluster, we applied \textit{data parallelism} at the loop level for all the parallelized components (i.e., QR decomposition, PSD, Vector coefficient update, Noise variance estimation, and Data preparation) for both AR and ARMA. This approach consists of partitioning the workload (associated with a loop index space) into multiple subsets assigned to the available cores.
Performing workflow partitioning implies an inherent overhead compared to the sequential code; to reduce this detrimental effect, we pre-compute the block size, start index, and end index for each core at the beginning of the algorithm by dividing the number of loop iterations by the number of cores.

\begin{figure}[t!]
\centerline{\includegraphics[width=0.45\textwidth]{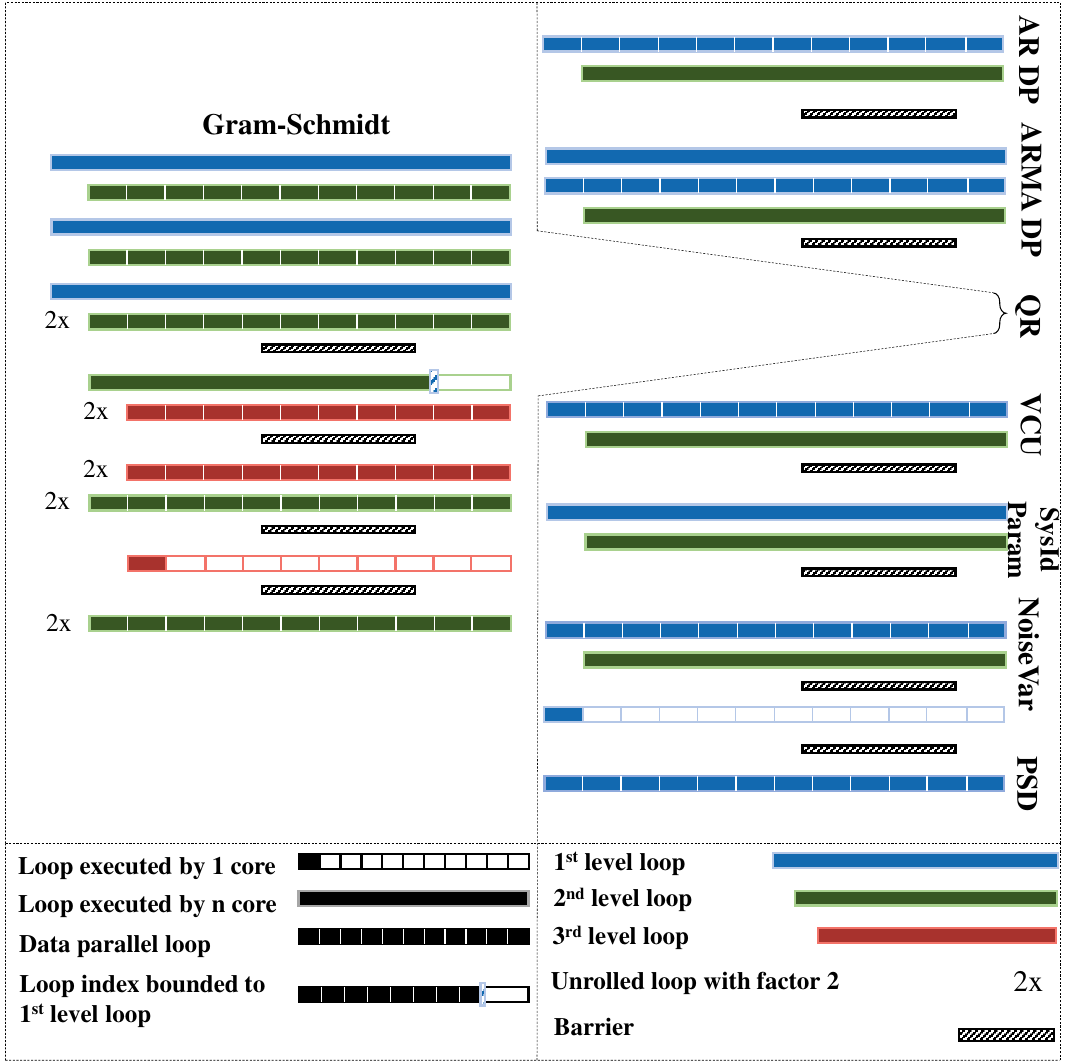}}
\caption{PARSY-VDD schematic structure on GAP9. Abbreviations: 'DP': data preparation (construction of the regression matrix). 'VCU': vector coefficient update procedure (i.e., $Q^T S$), 'SysIdParam': estimation of the model parameters $\Theta$, and 'noiseVar': noise variance estimation.}
\label{PARSY_Scheme}
\end{figure}

Fig.~\ref{PARSY_Scheme} illustrates the control flow of the different components of PARSY-VDD, providing insight into the parallelization approach. The data preparation step includes AR and ARMA; in each round, only one of these processes is utilized.
Each continuous bar represents a loop executed by all cores, the segmented bars depict loop iterations partitioned among multiple cores, and the black dashed bars denote synchronization points using \emph{barriers}.
The barrier construct ensures that all participating cores have executed the preceding instructions before proceeding to the following ones. As discussed in Sec.~\ref{subsec_gap9_theory}, the event unit in the GAP9 cluster enables synchronization constructs with minimum overhead; this component further reduces the system-level power consumption when the workload is not balanced, a common condition in signal processing algorithms.
Fig.~\ref{PARSY_Scheme} highlights two different synchronization patterns.
First, a barrier between two loops split among multiple cores guarantees data consistency.
Second, two barriers enclosing a loop executed by a single core guarantee the correctness of a reduction operation.

Most parallelizable loops (AR, ARMS, VCU, NoiseVar, PSD) are characterized by one or two nested levels, and parallelism is applied at the inner level to avoid data dependencies.
GS is the most complex kernel in the processing pipeline, including multiple nested loops.
The first two loops initialize the $Q$ and $R$ matrices to zero.
The last loop is deeply nested and transforms the elements under a pivot placed in the main diagonal; consequently, the number of loop iterations linearly decreases with the upper loop index. In this case, we define this loop as \emph{bounded} to the previous one.
This pattern could impact parallelization performance since the iteration space becomes too narrow to support workload partitioning on multiple cores (i.e., in all the cases where the iterations are less than 8).
However, we can perform parallelization on the third loop level, avoiding this issue by construction.
The following second-level loop performs a reduction. In this context, an alternative solution is a reduction tree that involves multiple cores. Our experiments assessed that single-core execution minimizes energy and execution time due to the high cost required to achieve mutex semantics on a shared data structure.

\subsubsection{Code-level optimizations}
To further optimize the execution time of our model, we applied two generic software optimizations at the C code level: loop unrolling and data transposition.
GAP9 already provides hardware loops and post-increment load/store operations to reduce the overhead of iterative code execution.
However, we used loop unrolling to reduce the number of pipeline stalls due to data dependencies.
To reduce \lone contentions when the number of row elements is a multiple of \qty{16} words (i.e., the number of banks), we applied transposition to invert the row-major layout of $C$ multidimensional arrays.

The process of calculating complex exponentials (see Eq. \ref{eq:psdarma} and \ref{eq:psdar}) is a critical aspect of the PSD kernel and requires a specific optimization. As a first step, we used the equivalent trigonometric expression for complex numbers since it allows the exploitation of sine and cosine values. However, invoking library functions requires over a hundred cycles. We employed a look-up table method for sine and cosine values to render the PSD implementation more energy and computation-efficient. This technique involves sampling the cosine values over a quarter period ($0$ to $\pi/2$) and then leveraging this table to determine the values for the full period of sine and cosine by simply performing a phase shift. This shift equates to reading a different index within the look-up table. Since the PSD sample count is fixed at 2048, this method incurs a memory overhead of \SI{2}{\kilo \byte} without any loss of accuracy.

\section{Performance profiling and comparison} \label{sec:hardware_res}

\begin{figure}[t!]
\centerline{\includegraphics[width=0.5\textwidth]{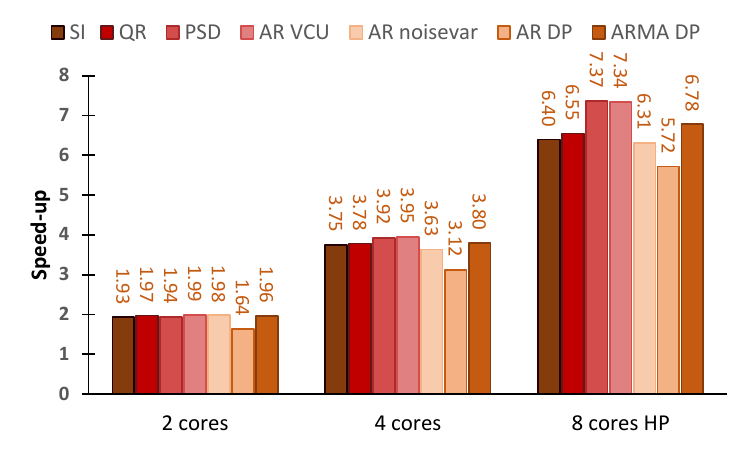}}
\caption{Speed-Up analysis of PARSY-VDD components and overall system across 2, 4, and 8 core HP configurations.}
\label{figure:SIDist-Speedup}
\end{figure}

\subsection{Experimental setup} 


\textcolor{black}{We conducted experiments on a GAP9 development board, which operates at a board voltage of \SI{1.8}{\volt}, using two configurations: (i) a high-performance (HP) setup running at \SI{370}{\mega\hertz} with a core voltage of \SI{0.8}{\volt}, tested with varying numbers of cores, and (ii) a low-power (LP) setup running at \SI{240}{\mega\hertz} with a core voltage of \SI{0.65}{\volt}, tested exclusively with 8 cores.}
To measure the power consumption, we utilized the Power Profiler Kit 2 (PPK2) from Nordic Semiconductor at a sampling rate of \SI{100}{\kilo sample/s}.
Each GAP9 core is equipped with a set of performance counters, including the Total Number of Cycles, the Total Number of Cycles Spent in Active Mode, the Number of Executed Instructions, Cycles Devoted to Load Data Hazard or Load Stalls, Instruction Cache Misses, the Number of External Loads, and Number of L1 data Contentions~\cite{GAP9}.
The SysId parameters have been configured as $N_p = 16$ and $N = 30N_p$ (necessary to fit the entire processing in the \lone memory of GAP9), keeping $L = 2048$ 32-bit precision data for the PSD, and selecting GS as QR method. Importantly, our analysis concentrates solely on the ARMA case in the subsequent sections. This choice is due to its superior spectral performance and higher complexity, which inherently encompasses AR, making it a comprehensive example for our comparisons.

\subsection{Performance and Energy Analysis}

\begin{table}[t]
\centering
\caption{\color{black}Instructions per cycle (IPC) for different platforms.}
\begin{tabular}{c | c c c c c c} 
 \hline
 & \textbf{F4} & \textbf{H7} & \textbf{GAP9 1} & \textbf{GAP9 2} & \textbf{GAP9 4} & \textbf{GAP9 8} \\ [0.5ex]  
 \hline
 \hline
IPC & 0.76 & 0.90 & 0.82 & 1.67 & 3.20 & 5.71\\
 \hline 
\end{tabular}
\captionsetup{width=0.5\linewidth}
\label{table:IPC}
\end{table}

\tabref{table:perf counter} reports a performance analysis of the implemented parallel SysId, including total execution cycles, percentage of cycles spent in different activities, memory-related statistics, and synchronization events (barriers). Barriers represent the number of synchronization events, while all other metrics are expressed as percentages of the total number of cycles.

By increasing the number of cores, the number of executed instructions per core decreases, reducing the cumulative instruction misses. However, this also increases the number of simultaneous L1 accesses, causing more conflicts and higher L1 data contention. The number of synchronization barriers remains relatively constant. In general, the increased L1 data contention and the overhead deriving from synchronization barriers are the main factors that prevent us from achieving the theoretical speed-up.

To gain a deeper understanding of the obtained performance enhancements, \textcolor{black}{Fig.}~\ref{figure:SIDist-Speedup} illustrates the speed-up achieved across various PARSY-VDD components, as well as the overall system speed-up.

The QR decomposition represents the largest component of the entire system. As we increase the size of the rows in the input matrix, the computation required for each core increases. This increase in computation helps to cover the overhead of parallelization, bringing the speed-up closer to the ideal value. However, despite this improvement, numerous synchronization barriers within the QR decomposition block prevent the capability to achieve the full performance gain. Indeed, according to the data shown in \textcolor{black}{Table}~\ref{table:perf counter}, out of 351 synchronization barriers, 336 are found within the QR decomposition block. 
Furthermore, another significant factor contributing to performance degradation is the necessity for sequential execution of square root ($SQRT$) and division operations required for vector norm calculation. Since the other parts of the code have been parallelized effectively and constitute a smaller portion of the total code, the overall speed-up of the entire system is ultimately bounded by the speed-up achieved in the QR decomposition. On the other hand, representing the second largest component of the PARSY-VDD, techniques such as sampling sine and cosine functions and utilizing a lookup table enhance the parallelization and optimization of the PSD. Consequently, the PSD component achieves a nearly ideal speed-up.

\begin{figure}[t!]
\centerline{\includegraphics[width=0.5\textwidth]{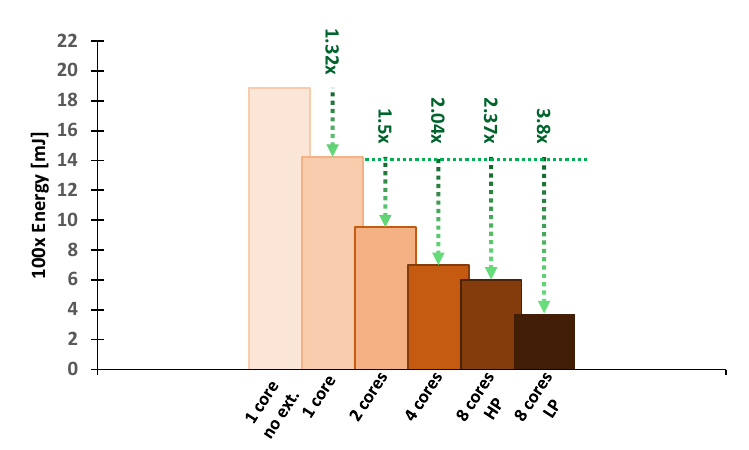}}
\caption{\textcolor{black}{Board energy consumption of PARSY-VDD deployed on GAP9 across different core configurations, including comparison with and without DSP instructions.}}
\label{figure:SIGap9-energy}
\end{figure}

\textcolor{black}{The parallelization and increase in speed lead to shorter execution times and, as shown in {Fig.}~\ref{figure:SIGap9-energy}, this also results in lower energy consumption. By comparing the single-core GAP9 setup without DSP instructions with the setup using them, we observe a reduction in energy consumption by 1.32×. This energy saving is due to using hardware loops, post-increment load/store instructions, and load/store with register offset for pointer arithmetic. Comparing the single-core setup on GAP9 with the HP 8-core configuration, using multiple cores not only speeds up execution but also results in an energy saving of 2.37×. Furthermore, by utilizing near-threshold computing and reducing the core voltage from 0.8 V to 0.65 V in the LP 8-core configuration, we can achieve an additional 1.6× reduction in energy consumption, bringing the total energy saving to 3.8× compared to sequential execution.}

\subsection{Comparison with commercial microcontrollers}

We extend our investigation to compare our design's performance and energy consumption with conventional single-core IoT devices, namely the high-end STM32H7 MCU for performance and the low-end STM32F4 MCU for energy comparison. Specifically, the STM32H7 has a frequency of \SI{400}{MHz} and a power consumption of \SI{234}{\milli W}; the STM32F4 has a core frequency of \SI{80}{MHz} and a power consumption of \SI{42.5}{\milli W}.

\subsubsection{Performance}

\begin{figure}[t!]
\centerline{\includegraphics[width=0.5\textwidth]{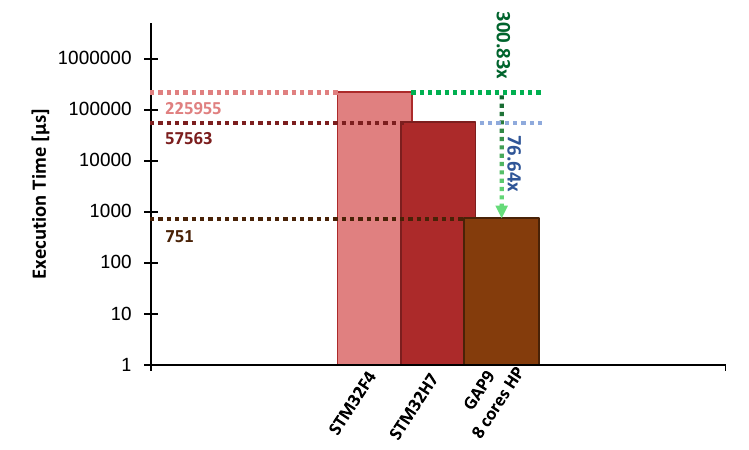}}
\caption{\textcolor{black}{Execution time of PARSY-VDD deployed on different low-end devices.}}
\label{figure:SI-performance}
\end{figure}

Fig.~\ref{figure:SI-performance} shows the excitation time and speed-up of the entire PARSY-VDD method across different platforms. \textcolor{black}{We observe performance gains of 76× using 8 cores HP configuration compared to STM32H7 as a high-performance single-core node}. Delving into the underlying reasons for these gains, we examined the assembly-generated code for insights. 

The analysis revealed that three primary factors contribute to the improved performance observed on the GAP9 architecture.
First, latency for all floating-point instructions (except division) remains constant at one cycle in GAP9~\cite{rossi2021vega}.
In contrast, the STM32H7 incurs a 3-cycle latency for floating-point operations, even if pipelining (when supported) can partially alleviate this issue. Second, the GAP9 architecture supports hardware loops that are unavailable in the STM32H7. Hardware loops allow the execution of a code segment multiple times without incurring branch overhead or counter updates, resulting in zero stall cycles during loop initiation. Finally, the last contributor to the performance gap is utilizing GAP9's post-increment load and store instructions. While STM32H7 architecture supports VL/SDMIA/DB instructions and zero overhead loops, the difference in implementation and latency also contributes to the observed gains.
As a final recap, \tabref{table:IPC} reports the instructions per cycle (IPC) achieved by each platform. On average, GAP9 has a higher IPC than STM32 platforms, even considering a single core. Moving to eight cores, the IPC scales less than linearly due to the overheads described in \tabref{table:perf counter}, but it guarantees a significant improvement.

\subsubsection{Energy efficiency}

\begin{figure}[t!]
\centerline{\includegraphics[width=0.5\textwidth]{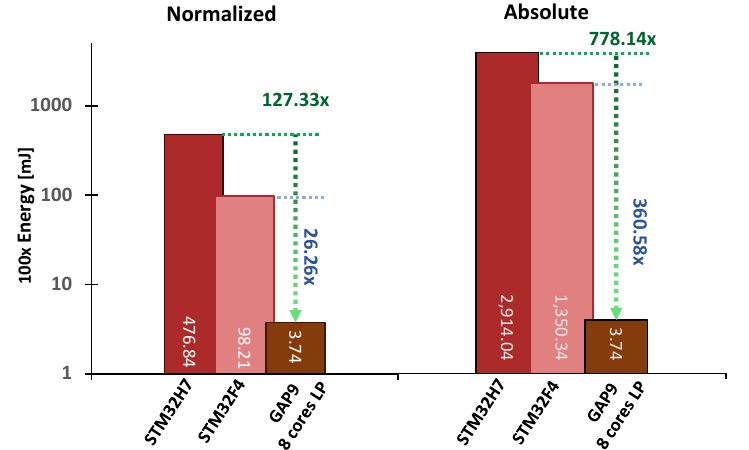}}
\caption{\textcolor{black}{Board energy consumption of PARSY-VDD deployed on different low-end devices. The left bars are the energy consumption normalized to \SI{22}{\nano m} technology of GAP9}.}
\label{figure:SI-Energy}
\end{figure}

~\figref{figure:SI-Energy} compares the actual and normalized energy consumption of PARSY-VDD in different platforms. 
The energy measurements are attained by executing each algorithm 100 times and then averaging the results. The technologies employed for GAP9, STM32H7, and STM32F4 are \SI{22}{\nano m}, \SI{40}{\nano m}, and \SI{90}{\nano m}, correspondingly. The related supply voltages for these technologies are \textcolor{black}{\SI{1.8}{V}, and \SI{3.3}{V}}. To ensure a fair assessment between the GAP9 and the STM32 series, the energy consumption values in \figref{figure:SI-Energy} are normalized to the \SI{22}{\nano m} scale using the power scaling factor $C \cdot V^2$. Here, $C$ and $V$ denote the effective capacitance (approximated with the channel length of the technology) and the supply voltage of the designs, respectively.

The near-threshold computing harnessed results in energy savings of \textcolor{black}{26.26×} for 8-core LP execution compared to STM32F4 execution as a low-power single-core node.

\section{Comparison with State-of the-Art} \label{sec:stoa}
\begin{table*}[h!]
\centering

\caption{System Identification @Edge: This table provides a comparison between several implementations of SysId in three different platforms: i) FPGA, ii) Single-Core MCU, iii) Multi-core MCU. Acronyms: RLS(Recursive Least Squares), NN (Neural Network), SHM (Structural Health Monitoring), PTS-QR (Parallel Tall Skinny - QR).}
\begin{adjustbox}{width=1\textwidth}
\label{tab:related_work}
\renewcommand{\arraystretch}{1.2}
\begin{tabular}{c|llc|lllll}
                                            & \textbf{Platform}          & \textbf{Application} & \textbf{End-to-End} & \textbf{Exec. Time}                  & \textbf{Power}           & \textcolor{black}{\textbf{Energy}} & \textbf{Memory}  & \textbf{SysId Kernel}                                                   \\ \hline

\multicolumn{8}{l}{\textcolor{black}{\textbf{FPGA-Based Implemnentations}} }   \\\hline
Atencia \textit{et al.}~\cite{atencia2007fpga}      & Spartan3 Family   & Robotic systems        & NO & N.A.                        & N.A.                   & \textcolor{black}{N.A.} & N.A.               & Hopfield Networks (NN)   \\\hline
Biradar \textit{et al.}~\cite{biradar2015fpga}      & Virtex-5 FPGA     & Motor Control          & NO & N.A.                        & \qty{400}{-}\qty{900}{\milli W}                    & \textcolor{black}{N.A.}  & N.A.               & Gradient-Based (NN)    \\\hline
KAPow ~\cite{hung2016kapow}                         & Altera Cyclone V  & Power Estimation       & NO & \qty{0.6}{\milli \second}   & \qty{1.8}{W}    & \textcolor{black}{\SI{1.08}{\milli J}}  & N.A.              & RLS      \\\hline
SysIdLib ~\cite{akgun2020sysidlib}                  & Zynq PYNQ-Z1      & DC Motor               & NO & \qty{750 }{\second}         & N.A.                   & \textcolor{black}{N.A.} & N.A.               & RLS      \\\hline
\multicolumn{8}{l}{\textbf{Single-Core Devices}}    \\\hline
Kim \textit{et al.}~\cite{kim2012autonomous}        &  ATmega128        & Data Reduction (SHM)   & NO & N.A.                        & \qty{85}{\milli W}     & \textcolor{black}{N.A.} & \qty{128}{\kilo B}        & Markov Parameter    \\\hline
Zonzini \textit{et al.}~\cite{zonzini2022system}    &  STM32L5          & Data Reduction (SHM)   & NO & \qty{129}{\second}          & \qty{49.5}{\milli W}   & \textcolor{black}{\SI{6.39}{J}} & \qty{256}{\kilo B}        & QR(Householder)    \\\hline
\multicolumn{8}{l}{\textbf{Multi-core Devices}}    \\\hline
Moallemi \textit{et al.}~\cite{moallemi2023speeding}& GAP9              & Data Reduction (SHM)   & NO & \textcolor{black}{\qty{68.10}{\milli \second}}  &       \textcolor{black}{\qty{47.13}{\milli W}}   & \textcolor{black}{\SI{3.21}{\milli J}} & \textcolor{black}{\qty{86.58}{\kilo B}}        & QR (Gram-Schmit)    \\ 
\hline  \hline 
\textbf{Our Work}                                   & GAP9              & Damage Detection (SHM) & Yes & \qty{0.75}{\milli \second}  &  \textcolor{black}{\qty{32.29}{\milli W}}  & \textcolor{black}{\SI{37.40}{\micro J}} & \qty{64.44}{\kilo B} & QR (Gram-Schmit)    \\\hline
\end{tabular}
  \end{adjustbox}
\end{table*}

Implementing SysId algorithms involves computational and memory-intensive operations, which justify the lack of experimental evidence for embedding SysId on tiny devices near the sensors. 
%
%
The first trial was made by Hann \textit{et al.} in ~\cite{hann2009real} by providing a C language script of a SysId library, while no deployment on edge devices was demonstrated.     
Field Programmable Gate Arrays (FPGA) gained prominence as a feasible option for real-time implementations of SysId due to their hardware-software co-design accessibility~\cite{atencia2007fpga, biradar2015fpga}. For instance, in~\cite{atencia2007fpga} and \cite{biradar2015fpga}, authors implement a hardware module that performs parametric identification of dynamical systems based on neural networks.   
%
Further, in~\cite{hung2016kapow}, SysId is solved via recursive least squares on an Altera Cyclone V device for real-time execution.
Lately, G{\"o}khan \textit{et al.}~\cite{akgun2020sysidlib} introduced SysIdLib, a high-level synthesis library for online SysId 
based on the Xilinx PYNQ-Z1 FPGA. 
%
%
\textcolor{black}{However, deploying FPGA-based systems at the edge is not intuitive due to their energy-demanding nature (more than $\approx \qty{1}{W}$), as they are primarily grid-based systems and consume an order of magnitude more energy than the commercially available low-end MCUs.
Further, the former FPGA-based approaches are based on various hardware architectures, making them incomparable to the fixed ASIC designs, i.e., microprocessors.  
}

{\tabref{tab:related_work} provides a summary of the embedded SysId algorithm across different platforms for data reduction and feature extraction across different applications like robotics and SHM several platforms.}
The first relevant example is provided in~\cite{kim2012autonomous}, which showcased SysId on the Narada sensor platform, i.e., one of the pristine and widely employed acceleration sensors for SHM based on the Atmel ATmega128 MCU operating at \SI{8}{MHz} clock frequency and maximum storage capabilities of \SI{128}{kBytes} of RAM. Their implementation consumes \qty{85}{\milli W}, which is the average power consumption of the Atmel ATmega128 in the fully operating mode where no duty cycling technique is used during acquisition and computing. However, the hardware/software constraints of the system imposed the adoption of correlation-based SysId variants, hence hampering its effective exploitation as an efficient data reduction method.  
%
%

%
Recently, authors in~\cite{zonzini2022system} achieved successful deployment on a general-purpose prototype sensor equipped with an STM32L5 MCU; however, this solution is affected by long execution time (greater than \SI{120}{s} in the most cumbersome scenario) due to the sequential nature of the computational workflow forced by the single-core architecture of the computing processor. 
%

%
%
\textcolor{black}{More recently, Moallemi \textit{et al.}~\cite{moallemi2023speeding} introduced a preliminary study of the feasibility of implementing parallel QR for SysId on a multi-core device to decrease the latency of the method and mitigate the data rate to the cloud. Their parallelization approach is based on a chunking scheme that applies parallel QR on adjacent data blocks and then recombines intermediate results until the sought $Q$ and $R$ matrices are extracted according to a binary decomposition tree. Thanks to this paradigm, authors were able to reduce the energy consumption by two orders of magnitude compared with \cite{zonzini2022system}.}
%

\textcolor{black}{Conversely, while exploiting the same MCU, our work substantially differentiates from the previous attempt in \cite{moallemi2023speeding} in the following relevant aspects: i) we present a fully optimized end-to-end damage detection processing flow including not only SysId (which was the main contribution of \cite{moallemi2023speeding}) but also the computation of the PSD necessary for damage detection, along with the data preparation and feature extraction process that might have a huge impact on the overall computational complexity, ii) we adopt a different parallelization model for the QR factorization to properly exploit the GAP9 multi-core capabilities. }
%
%
\textcolor{black}{To this end, we have made two main modifications. First, we have removed the data chunking and processed the entire input signal in one shot of QR decomposition, hence removing the parallelization overheads of~\cite{moallemi2023speeding}, resulting in a speed-up of \textcolor{black}{\qty{90.8}{\times}} in latency and \textcolor{black}{85.8$\times$} in energy efficiency. Second, we have adopted an economy-size variant of the QR function, leading us to fit medium model orders, i.e., $Np = 16  \times N = 400$, of SysId in \lone memory with a memory footprint reduction of \textcolor{black}{\qty{18}{\%}} (in combination with the fact that intermediate QR results are not necessary to be stored in temporary memory.}
Finally, through an evaluative comparison of the PARSY-VDD framework across various single-core and multi-core edge devices, we show that the deployment of the algorithm on GAP9 facilitates the development of a long-lasting SHM node capable of transmitting minimal data volumes to the cloud.
Consequently, our methodology offers a framework for a dense, scalable monitoring infrastructure paving the way to large-scale structural assessments at a country level.
\section{Conclusion} \label{sec:conc}


This paper introduces PARSY-VDD, a parallelized end-to-end framework designed for near-sensor vibration diagnosis using SysId algorithms. PARSY-VDD addresses key challenges at the extreme edge, such as high latency and power consumption, while preserving structural information. Evaluation of two significant structural health monitoring (SHM) datasets, i.e., the Z24 bridge and a wind turbine blade, demonstrates its effectiveness.

%
%

From a structural viewpoint, this has been demonstrated by assessing the detection capabilities of PARSY-VDD dealing with two representative SHM datasets, i.e., the Z24 bridge and a wind turbine blade. A spectral-based approach has been used, which tracks variation in the location of the peak values of the PSD profiles obtained from SysId parameters between the healthy and the defective configurations. Firstly, the quality of these spectra with respect to a golden model running on a laptop PC has been measured via the ISD metric, scoring a worst-case value of $1.91\cdot10^{\text{-}4}$ which is very proximal to zero and, thus, it indicates very consistent matching. Then, we have shown that
%
PARSY-VDD is able to identify a percentage variation of \qty{2.59}{\%} and \qty{4.54}{\%} for the Z24 and wind turbine, respectively. 
%



From an architectural perspective, the performance speed-up of PARSY-VDD leverages the parallel ultra-low power capabilities of GAP9, a multi-core RISC-V device.
%
Indeed, by exploiting the unique multi-core architecture of GAP9 and its capability to operate at a near-threshold voltage, running PARSY-VDD on GAP9 reached a speed-up of \qty{76}{\times} in execution time and energy efficiency of \qty{360}{\times} compared to commercial single-core devices, i.e., STM32H7 and STM32F4 MCUs, respectively. \textcolor{black}{Noteworthy, thanks to advanced parallelization strategies, improvement with respect to the state-of-the-art, i.e., a previous work in which SysId alone was deployed on the same GAP9 platform without including the final PSD stage, resulted in a speed-up of \qty{90}{\times} in latency and \qty{85}{\times} in energy efficiency.}

\bibliographystyle{IEEEtran}
\bibliography{bibliography.bib}

\begin{thebibliography}{10}
\providecommand{\url}[1]{#1}
\csname url@samestyle\endcsname
\providecommand{\newblock}{\relax}
\providecommand{\bibinfo}[2]{#2}
\providecommand{\BIBentrySTDinterwordspacing}{\spaceskip=0pt\relax}
\providecommand{\BIBentryALTinterwordstretchfactor}{4}
\providecommand{\BIBentryALTinterwordspacing}{\spaceskip=\fontdimen2\font plus
\BIBentryALTinterwordstretchfactor\fontdimen3\font minus \fontdimen4\font\relax}
\providecommand{\BIBforeignlanguage}[2]{{%
\expandafter\ifx\csname l@#1\endcsname\relax
\typeout{** WARNING: IEEEtran.bst: No hyphenation pattern has been}%
\typeout{** loaded for the language `#1'. Using the pattern for}%
\typeout{** the default language instead.}%
\else
\language=\csname l@#1\endcsname
\fi
#2}}
\providecommand{\BIBdecl}{\relax}
\BIBdecl

\bibitem{giordano2023value}
P.~F. Giordano, S.~Quqa, and M.~P. Limongelli, ``The value of monitoring a structural health monitoring system,'' \emph{Structural Safety}, vol. 100, p. 102280, 2023.

\bibitem{iasha2020design}
F.~Iasha and P.~A. Darwito, ``Design of algorithm control for monitoring system and control bridge based internet of things (iot),'' in \emph{2020 International Conference on Smart Technology and Applications (ICoSTA)}.\hskip 1em plus 0.5em minus 0.4em\relax IEEE, 2020, pp. 1--6.

\bibitem{sohn2001damage}
H.~Sohn and C.~R. Farrar, ``Damage diagnosis using time series analysis of vibration signals,'' \emph{Smart materials and structures}, vol.~10, no.~3, p. 446, 2001.

\bibitem{morgese2020post}
M.~Morgese, F.~Ansari, M.~Domaneschi, and G.~P. Cimellaro, ``Post-collapse analysis of morandi’s polcevera viaduct in genoa italy,'' \emph{Journal of Civil Structural Health Monitoring}, vol.~10, pp. 69--85, 2020.

\bibitem{figueiredo2013linear}
E.~Figueiredo and E.~Cross, ``Linear approaches to modeling nonlinearities in long-term monitoring of bridges,'' \emph{Journal of Civil Structural Health Monitoring}, vol.~3, pp. 187--194, 2013.

\bibitem{dipietrangelo2023structural}
F.~Dipietrangelo, F.~Nicassio, and G.~Scarselli, ``Structural health monitoring for impact localisation via machine learning,'' \emph{Mechanical Systems and Signal Processing}, vol. 183, p. 109621, 2023.

\bibitem{parisi2022time}
E.~Parisi, A.~Moallemi, F.~Barchi, A.~Bartolini, D.~Brunelli, N.~Buratti, and A.~Acquaviva, ``Time and frequency domain assessment of low-power mems accelerometers for structural health monitoring,'' in \emph{2022 IEEE International Workshop on Metrology for Industry 4.0 \& IoT (MetroInd4. 0\&IoT)}.\hskip 1em plus 0.5em minus 0.4em\relax IEEE, 2022, pp. 234--239.

\bibitem{cai2016iot}
H.~Cai, B.~Xu, L.~Jiang, and A.~V. Vasilakos, ``Iot-based big data storage systems in cloud computing: perspectives and challenges,'' \emph{IEEE Internet of Things Journal}, vol.~4, no.~1, pp. 75--87, 2016.

\bibitem{saidin2023vibration}
S.~S. Saidin, S.~A. Kudus, A.~Jamadin, M.~A. Anuar, N.~M. Amin, A.~B.~Z. Ya, and K.~Sugiura, ``Vibration-based approach for structural health monitoring of ultra-high-performance concrete bridge,'' \emph{Case Studies in Construction Materials}, vol.~18, p. e01752, 2023.

\bibitem{kamariotis2023framework}
A.~Kamariotis, E.~Chatzi, and D.~Straub, ``A framework for quantifying the value of vibration-based structural health monitoring,'' \emph{Mechanical Systems and Signal Processing}, vol. 184, p. 109708, 2023.

\bibitem{polonelli2023self}
T.~Polonelli, A.~Moallemi, W.~Kong, H.~M{\"u}ller, J.~Deparday, M.~Magno, and L.~Benini, ``A self-sustainable and micro-second time synchronized multi-node wireless system for aerodynamic monitoring on wind turbines,'' \emph{IEEE Access}, vol.~11, pp. 119\,506--119\,522, 2023.

\bibitem{di2021structural}
F.~Di~Nuzzo, D.~Brunelli, T.~Polonelli, and L.~Benini, ``Structural health monitoring system with narrowband iot and mems sensors,'' \emph{IEEE Sensors Journal}, vol.~21, no.~14, pp. 16\,371--16\,380, 2021.

\bibitem{ballerini2020nb}
M.~Ballerini, T.~Polonelli, D.~Brunelli, M.~Magno, and L.~Benini, ``Nb-iot versus lorawan: An experimental evaluation for industrial applications,'' \emph{IEEE Transactions on Industrial Informatics}, vol.~16, no.~12, pp. 7802--7811, 2020.

\bibitem{zonzini2022system}
F.~Zonzini, V.~Dertimanis, E.~Chatzi, and L.~De~Marchi, ``System identification at the extreme edge for network load reduction in vibration-based monitoring,'' \emph{IEEE Internet of Things Journal}, vol.~9, no.~20, pp. 20\,467--20\,478, 2022.

\bibitem{moallemi2023speeding}
A.~Moallemi, R.~Gaspari, F.~Zonzini, L.~De~Marchi, D.~Brunelli, and L.~Benini, ``Speeding up system identification algorithms on a parallel risc-v mcu for fast near-sensor vibration diagnostic,'' \emph{IEEE Sensors Letters}, 2023.

\bibitem{GAP9}
\BIBentryALTinterwordspacing
G.~Technologies. (2014) {GreenWaves Technologies} gap9 official description. [Online]. Available: \url{{https://greenwaves-technologies.com/gap9$\_$processor/}}
\BIBentrySTDinterwordspacing

\bibitem{rossi2021vega}
D.~Rossi, F.~Conti, M.~Eggiman, A.~Di~Mauro, G.~Tagliavini, S.~Mach, M.~Guermandi, A.~Pullini, I.~Loi, J.~Chen \emph{et~al.}, ``Vega: A ten-core soc for iot endnodes with dnn acceleration and cognitive wake-up from mram-based state-retentive sleep mode,'' \emph{IEEE Journal of Solid-State Circuits}, vol.~57, no.~1, pp. 127--139, 2021.

\bibitem{jahangiri2023procedure}
M.~Jahangiri, A.~Palermo, S.~Kamali, M.~A. Hadianfard, and A.~Marzani, ``A procedure to estimate the minimum observable damage in truss structures using vibration-based structural health monitoring systems,'' \emph{Probabilistic Engineering Mechanics}, vol.~73, p. 103451, 2023.

\bibitem{zonzini2023tiny}
F.~Zonzini, L.~Burioli, A.~Gashi, N.~F. Mancini, and L.~De~Marchi, ``A tiny convolutional neural network driven by system identification for vibration anomaly detection at the extreme edge,'' in \emph{2023 IEEE International Conference on Omni-layer Intelligent Systems (COINS)}.\hskip 1em plus 0.5em minus 0.4em\relax IEEE, 2023, pp. 1--6.

\bibitem{van1996matrix}
C.~F. Van~Loan and G.~Golub, ``Matrix computations (johns hopkins studies in mathematical sciences),'' \emph{Matrix Computations}, vol.~5, 1996.

\bibitem{GolubVanLoan2012}
G.~H. Golub and C.~F.~V. Loan, \emph{Matrix Computations}, 4th~ed.\hskip 1em plus 0.5em minus 0.4em\relax Johns Hopkins University Press, 2012.

\bibitem{yiu2013definitive}
J.~Yiu, \emph{The Definitive Guide to ARM{\textregistered} Cortex{\textregistered}-M3 and Cortex{\textregistered}-M4 Processors}.\hskip 1em plus 0.5em minus 0.4em\relax Newnes, 2013.

\bibitem{Gautschi2017}
M.~Gautschi, P.~D. Schiavone, A.~Traber, I.~Loi, A.~Pullini, D.~Rossi, E.~Flamand, F.~K. Gürkaynak, and L.~Benini, ``Near-threshold risc-v core with dsp extensions for scalable iot endpoint devices,'' \emph{IEEE Transactions on Very Large Scale Integration (VLSI) Systems}, vol.~25, no.~10, pp. 2700--2713, 2017.

\bibitem{ou2021vibration}
Y.~Ou, K.~E. Tatsis, V.~K. Dertimanis, M.~D. Spiridonakos, and E.~N. Chatzi, ``Vibration-based monitoring of a small-scale wind turbine blade under varying climate conditions. part i: An experimental benchmark,'' \emph{Structural Control and Health Monitoring}, vol.~28, no.~6, p. e2660, 2021.

\bibitem{10.1145/3649153.3649210}
\BIBentryALTinterwordspacing
A.~Kiamarzi, D.~Rossi, and G.~Tagliavini, ``Qr-pulp: Streamlining qr decomposition for risc-v parallel ultra-low-power platforms,'' in \emph{Proceedings of the 21st ACM International Conference on Computing Frontiers}, ser. CF '24.\hskip 1em plus 0.5em minus 0.4em\relax New York, NY, USA: Association for Computing Machinery, 2024, p. 147–154. [Online]. Available: \url{https://doi.org/10.1145/3649153.3649210}
\BIBentrySTDinterwordspacing

\bibitem{brincker2015introduction}
R.~Brincker and C.~Ventura, \emph{Introduction to operational modal analysis}.\hskip 1em plus 0.5em minus 0.4em\relax John Wiley \& Sons, 2015.

\bibitem{atencia2007fpga}
M.~Atencia, H.~Boumeridja, G.~Joya, F.~Garc{\'\i}a-Lagos, and F.~Sandoval, ``Fpga implementation of a systems identification module based upon hopfield networks,'' \emph{Neurocomputing}, vol.~70, no. 16-18, pp. 2828--2835, 2007.

\bibitem{biradar2015fpga}
R.~G. Biradar, A.~Chatterjee, K.~George, and P.~Mishra, ``Fpga implementation of learning for online system identification,'' in \emph{2015 International Conference on Computing and Network Communications (CoCoNet)}.\hskip 1em plus 0.5em minus 0.4em\relax IEEE, 2015, pp. 274--282.

\bibitem{hung2016kapow}
E.~Hung, J.~J. Davis, J.~M. Levine, E.~A. Stott, P.~Y. Cheung, and G.~A. Constantinides, ``Kapow: A system identification approach to online per-module power estimation in fpga designs,'' in \emph{2016 IEEE 24th Annual International Symposium on Field-Programmable Custom Computing Machines (FCCM)}.\hskip 1em plus 0.5em minus 0.4em\relax IEEE, 2016, pp. 56--63.

\bibitem{akgun2020sysidlib}
G.~Akg{\"u}n, H.~u.~H. Khan, M.~Hebaish, M.~Elshimy, M.~A. A.~E. Ghany, and D.~G{\"o}hringer, ``Sysidlib: A high-level synthesis fpga library for online system identification,'' in \emph{Applied Reconfigurable Computing. Architectures, Tools, and Applications: 16th International Symposium, ARC 2020, Toledo, Spain, April 1--3, 2020, Proceedings 16}.\hskip 1em plus 0.5em minus 0.4em\relax Springer, 2020, pp. 97--107.

\bibitem{kim2012autonomous}
J.~Kim and J.~P. Lynch, ``Autonomous decentralized system identification by markov parameter estimation using distributed smart wireless sensor networks,'' \emph{Journal of Engineering Mechanics}, vol. 138, no.~5, pp. 478--490, 2012.

\bibitem{hann2009real}
C.~E. Hann, I.~Singh-Levett, B.~L. Deam, J.~B. Mander, and J.~G. Chase, ``Real-time system identification of a nonlinear four-story steel frame structure—application to structural health monitoring,'' \emph{IEEE Sensors Journal}, vol.~9, no.~11, pp. 1339--1346, 2009.

\end{thebibliography}

\begin{IEEEbiography}[{\includegraphics[width=1in,height=1.25in,clip,keepaspectratio]{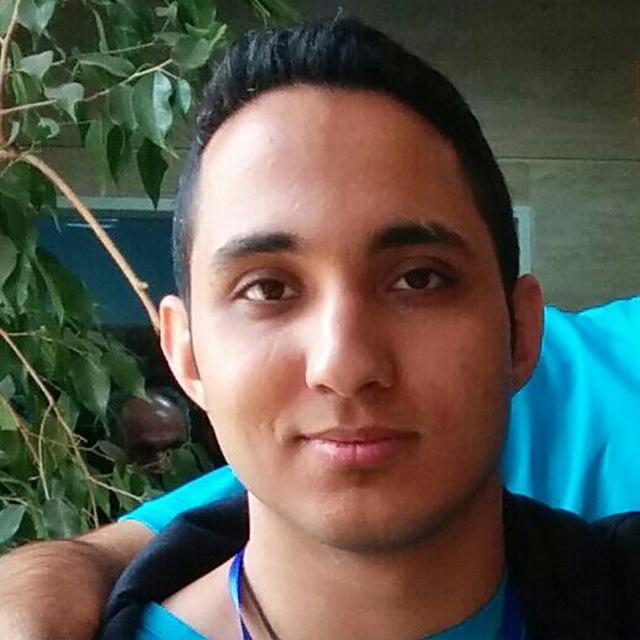}}]{Amirhossein Kiamarzi} received his Master’s degree in Computer Engineering from the University of Shiraz in 2020. He is currently pursuing a Ph.D. in Digital Systems Design at the Department of Electrical and Information Engineering (DEI), University of Bologna. His research focuses on modeling RISC-V-based hardware cores and accelerators.
\end{IEEEbiography}

\begin{IEEEbiography}[{\includegraphics[width=1in,height=1.25in,clip,keepaspectratio]{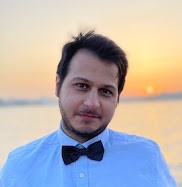}}]{Amirhossein Moallemi} received his M.Sc. (cum laude) and Ph.D. degree in electronics engineering from the University of Bologna, Bologna, Italy, in 2020 and 2024, respectively.
He is currently a senior electronics engineer at RTDT Laboratories, AG, Zurich, Switzerland. His research interests include IoT, low-power hardware and firmware embedded systems design, machine and deep learning models, and structural health monitoring systems.
\end{IEEEbiography}

\begin{IEEEbiography}[{\includegraphics[width=1in,height=1.25in,clip,keepaspectratio]{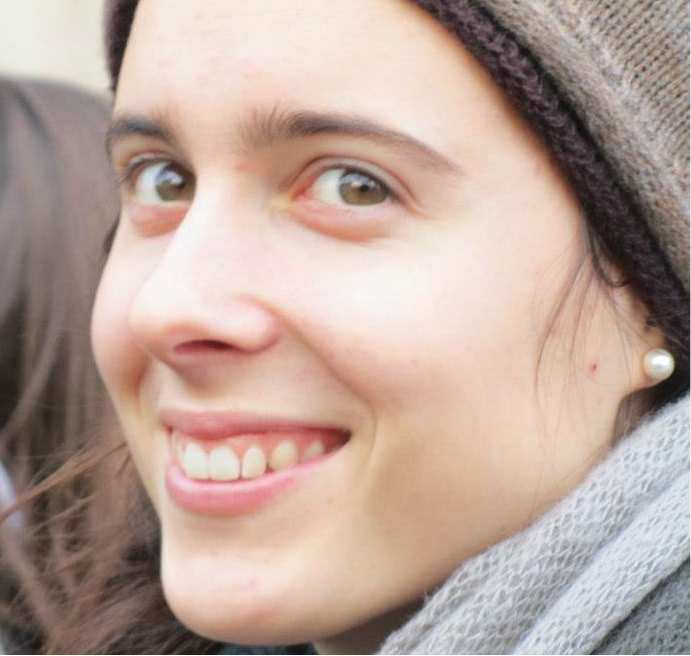}}]{Federica Zonzini} received the B.S. and M.S. degrees in Electronic Engineering and the Ph.D. degree in Structural and Environmental Health Monitoring and Management (SEHM2) from the University of Bologna, Bologna, Italy, in 2016, 2018, and 2022, respectively.
She is Junior Research Assistant in electronics with the University of Bologna, Bologna, Italy. Her research interests include the design of intelligent sensor systems and edge computing in the context of structural health monitoring, encompassing advanced signal processing, and tiny machine learning.
\end{IEEEbiography}

\begin{IEEEbiography}[{\includegraphics[width=1in,height=1.25in,clip,keepaspectratio]{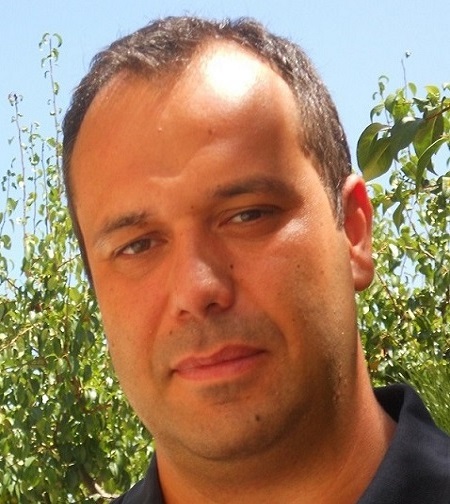}}]{Davide Brunelli} received the M.S. (cum laude) and Ph.D. degrees in electrical engineering from the University of Bologna, Italy, in 2002 and 2007, respectively. He is an Associate Professor of electronics with the Department of Industrial Engineering, University of Trento, Italy.
He has published more than 300 research papers in international conferences and journals on ultra-low-power embedded systems, energy harvesting, and power management of VLSI circuits.
He holds several patents and is annually ranked among the top 2\% of scientists according to the “Stanford World Ranking of Scientists” from 2020.
His current research interests include new techniques of energy scavenging for IoT and embedded systems, the optimization of low-power and low-cost consumer electronics, and the interaction and design issues in embedded personal and wearable devices. He is a member of several TPC conferences in the Internet of Things (IoT) and is an Associate Editor of the IEEE.
\end{IEEEbiography}

\begin{IEEEbiography}[{\includegraphics[width=1in,height=1.25in,clip,keepaspectratio]{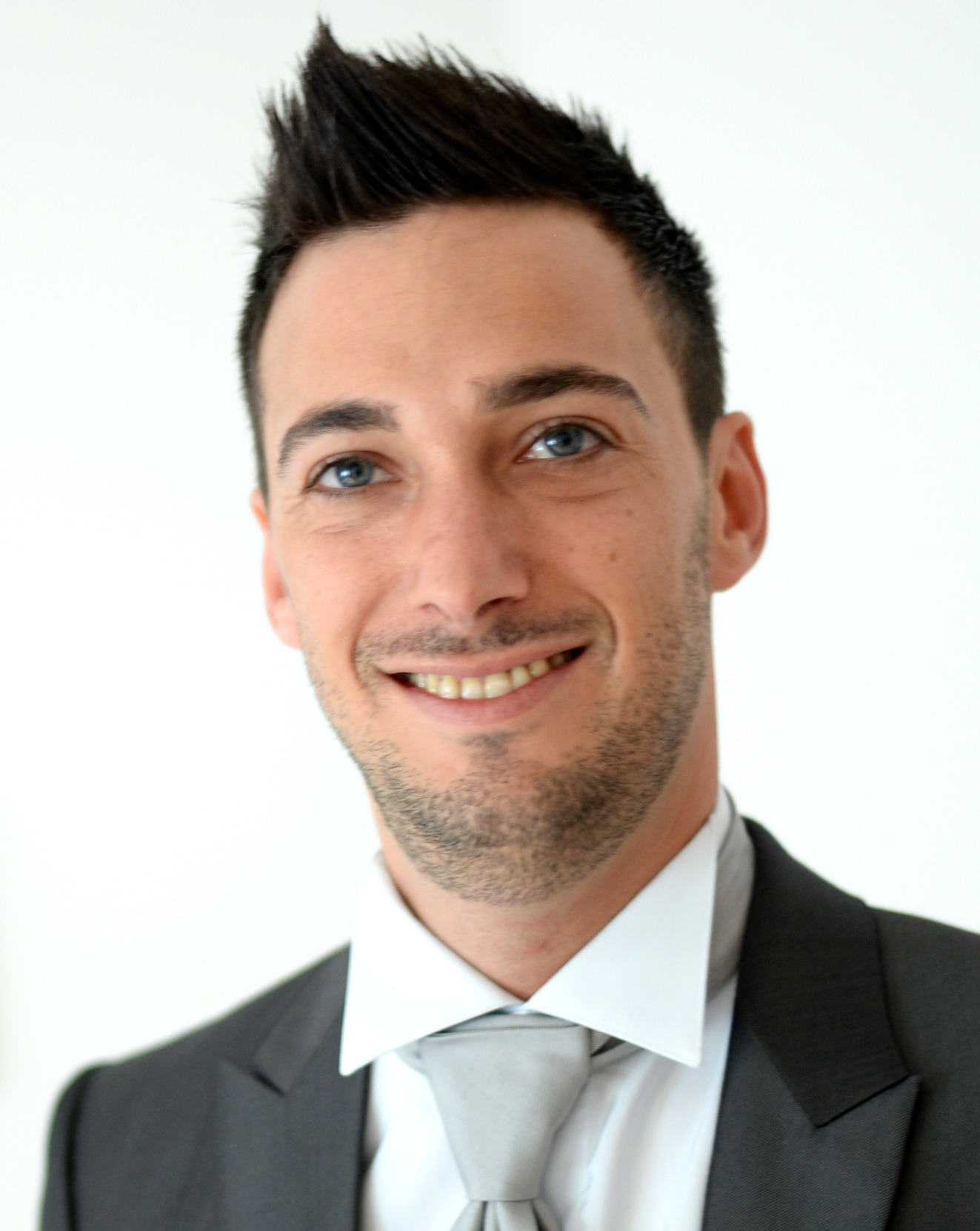}}]{Davide Rossi} received the Ph.D. degree from the University of Bologna, Bologna, Italy, in 2012. He has been a Post-Doctoral Researcher with the Department of Electrical, Electronic and Information Engineering “Guglielmo Marconi,” University of Bologna, since 2015, where he is currently an Associate Professor. His research interests focus on energy-efficient digital architectures. In this field, he has published more than 100 papers in international peer-reviewed conferences and journals. He is recipient of Donald O. Pederson Best Paper Award 2018, 2020 IEEE TCAS Darlington Best Paper Award, 2020 IEEE TVLSI Prize Paper Award.
\end{IEEEbiography}

\begin{IEEEbiography}[{\includegraphics[width=1in,height=1.25in,clip,keepaspectratio]{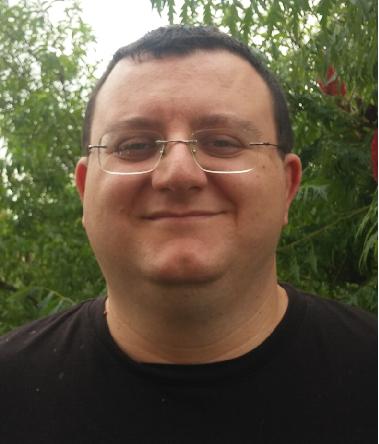}}]{Giuseppe Tagliavini} is a Tenure-Track Assistant Professor at the University of Bologna. His research interests are focused on programming models, orchestration tools, and compiler optimizations for AI-enabled resource-constrained computing platforms, with a strong emphasis on parallel architectures and accelerators in the context of ultra-low-power IoT end nodes. He is a member of the IEEE and ACM societies.
\end{IEEEbiography}

\end{document}